\documentclass[aps,showpacs,prb,twocolumn,superscriptaddress]{revtex4}
\usepackage{times,xspace}
\usepackage{amsbsy,amssymb,amsmath,bm}
\usepackage{graphicx,color,epsfig,rotate}
\usepackage{fancyheadings}

\begin{document}
\title{FERROELASTIC DYNAMICS AND STRAIN COMPATIBILITY}
\author{T. Lookman}
\affiliation{Theoretical Division and Center for Nonlinear Studies, Los 
Alamos
National Laboratory,
Los Alamos, New Mexico, 87545}
\author{S.~R. Shenoy}
\affiliation{Abdus Salam International Centre for Theoretical Physics, 34100
Trieste, Italy}
\author{K.~{\O}. Rasmussen}
\author{A. Saxena}
\author{A.R. Bishop}
\affiliation{Theoretical Division and Center for Nonlinear Studies, Los 
Alamos
National Laboratory,
Los Alamos, New Mexico, 87545}
\date{\today}



\begin{abstract}

We derive  underdamped evolution equations for the order-parameter
$(OP)$ strains  of a ferroelastic material undergoing a structural
transition, using Lagrangian variations with Rayleigh dissipation, and
a free energy as a polynomial expansion in the $N=n+N_{op}$ symmetry-adapted
strains. The $N_{op}$ strain equations are structurally similar in
form to the Lagrange-Rayleigh $1D$
strain dynamics of Bales and Gooding ($BG$), with `strain accelerations'
proportional to a Laplacian acting on a sum of the free energy strain
derivative and frictional strain force. The tensorial St. Venant's
elastic compatibility
constraints that forbid defects, are used to determine the $n$
non-order-parameter strains in terms of the $OP$ strains,
generating  anisotropic and long-range $OP$
contributions to the free energy, friction and noise.
The {\it same} $OP$
equations are obtained by either varying the displacement vector
components,
or by  varying the $N$ strains subject to the $N_c$ compatibility
constraints. A Fokker-Planck equation, based on the $BG$ dynamics with
noise terms, is set up.
The $BG$ dynamics corresponds to a set of nonidentical nonlinear
(strain) oscillators labeled by wavevector $\vec{k}$, with
competing short- and long-range couplings. The oscillators have different
`strain-mass' densities $\rho (k) \sim 1/k^2$ and dampings
$\sim 1/\rho (k) \sim k^2$, so the lighter large-$k$ oscillators
equilibrate first, corresponding to earlier formation of smaller-scale
oriented textures. This produces
a sequential-scale scenario for post-quench nucleation,
elastic patterning, and hierarchical growth.
Neglecting inertial effects yields a  late-time dynamics for identifying
extremal free energy states, that is of the
 time-dependent
Ginzburg-Landau  form, with  nonlocal, anisotropic Onsager
coefficients, that become constants  for special parameter values.
We consider in detail the two-dimensional ($2D$) unit-cell transitions from
a triangular to a centered rectangular lattice 
$(N_{op}=2, n=1, N_c=1)$; and from a square to a rectangular lattice
$(N_{op}=1, n=2, N_c=1)$ for which the $OP$ compatibility kernel is
retarded in time, or frequency-dependent in Fourier space (in fact,
acoustically resonant in $\omega/k$).  We
present structural dynamics  for all
other $2D$ symmetry-allowed ferroelastic transitions: the procedure is
also applicable to the $3D$ case.
Simulations  of the $BG$ evolution equations confirm the inherent richness
of the static and dynamic texturings, including
strain oscillations, domain-wall propagation at near sound speeds,
grain-boundary motion,
and nonlocal `elastic photocopying' of imposed local stress patterns.
\end{abstract}
\maketitle

\section{INTRODUCTION}

Structural phase transitions in solids have attracted a great deal of
interest over a century, both for their conceptual importance as
symmetry-changing phase transitions, and for their role in inducing
technologically useful materials properties. Both the diffusion-controlled
replacive and the diffusionless displacive  transformations
have been studied, although the former have received  more attention
because their reaction kinetics is  more conducive to
control.

We consider here the class of materials known as ferroelastic
martensites.
Ferroelasticity is defined by the existence of two or more stable
orientation states of a crystal that correspond to different
arrangements of the atoms, but are structurally identical or
enantiomorphous \cite{wadhawan,martens}. In addition, these orientation
states are degenerate in energy in the absence of mechanical stress.
The term `martensitic' refers to a diffusionless first
order phase transition which can be described in terms of one (or
several successive) shear deformation(s) from a parent to a product
phase \cite{nishi}.
The morphology
and kinetics of the transition are dominated by the strain energy.
The transition results in a characteristic
lamellar or twinned microstrcuture.

Salient features of ferroelastic crystals include mechanical hysteresis
and mechanically (reversibly) switchable domain patterns. Usually
ferroelasticity occurs as a result of a phase transition from a
non-ferroelastic high-symmetry `parent' phase and is associated
with the softening of an elastic modulus with decreasing temperature
or increasing pressure in the parent phase.
Since the ferroelastic
transition is normally weakly first order, or second order, it can be
described to a good approximation by the Landau theory with spontaneous
strain or deviation of a given ferroelastic orientation state from the
parent phase as the order parameter.
The strain can be coupled
to other fields such as electric polarization and magnetic moment and 
thus the crystal can have more than one transition.
Depending on whether the spontaneous
strain is the primary or a secondary order
parameter at a given transition, the lower
symmetry phase is called a proper or an improper
ferroelastic, respectively.  While martensites are proper ferroelastics,
examples of improper ferroelastics include ferroelectrics and
magnetoelastics.

There is a further subset of ferroelastic martensites (either non-elemental 
metals or alloy systems) that exhibit the shape memory
effect \cite{otsuka}. These materials are characterized by highly mobile twin
boundaries and (often) show precursor structures above the transition.
Furthermore, these materials have small Bain strain, elastic shear
modulus softening, and a weakly to moderately first order transition.
Some examples include InTl, FePd, NiTi and AuCd.

Dynamics plays a central role in proper ferroelastic transitions
\cite{martens,BG,Reid,Chen,Khatch,kerr,ourRC,our3DPRL,Yama,
Onuki,Anantha,Madan,JacobsHO,JacobsSR}.  As noted, these materials undergo
diffusionless, displacive transitions, with strain (components) as
the order parameter, and develop complex microstructures in their
dynamical evolution, finally forming spatially varying, multiscale
`textures' or strain patterns.  When quenched, some martensitic
materials develop interfaces moving at near-sound speeds.
Textured improper ferroelastics include materials of technological
importance such as superconducting cuprates \cite{YBCO} and colossal
magnetoresistance (CMR) manganites \cite{asamitsu}.  Many dynamical
models have been invoked to follow aspects of (proper) ferroelastic
pattern formation
\cite{BG,Reid,Chen,Khatch,kerr,ourRC,our3DPRL,Yama,Onuki,Anantha,Madan,
JacobsHO,JacobsSR} such as nucleated twin-front propagation;
width-length scaling of twin dimensions \cite{moore,bhk}; tweed
\cite{sugiyama,sb,Kartha}; stress effects; elastic domain misfits;
and acoustic noise generation \cite{planes}.

In a one-dimensional ($1D$) model, Bales and Gooding \cite{BG}
considered a displacement $u$ and a sixth order free energy
$F(\varepsilon )$, nonlinear in the
strain,
that in $1D$ is simply a derivative, $\varepsilon =\partial u/\partial x$.
  With
Lagrangian dynamics, a Rayleigh dissipation \cite{LL} and variation in
$u$,
a single strain evolution equation was obtained in $1D$:
$$\rho_0 \ddot{\varepsilon} = \frac{\partial^2}{\partial x^2}
\left( {\frac{\delta
F}{\delta\varepsilon}}+
A' \dot{\varepsilon}\right), \eqno(1.1)$$
where $A'$ and $\rho_0$ are the scaled friction (coefficient) and mass density,
respectively. In a low-frequency/large-wavevector regime $\rho_{0}\omega << A'k^2$,
where the inertial term is small, a simple time-dependent
Ginzburg-Landau $(TDGL)$ equation is obtained  \cite{BG,Reid} in $1D$:
$$\dot{\varepsilon} =-\frac{1}{A'}\frac {\delta F}{\delta\varepsilon}.
\eqno(1.2)$$
For the $k \ne 0$ modes in $1D$ the strain is a one-component vector that in Fourier space is
simply
proportional to the scalar displacement; and there
is only one type of lattice symmetry.
In higher dimensions $D>1$, the  strain (displacement)  is a tensor
(vector),
and there are many possible discrete lattice symmetries.
The central question is:
What is the general form of the underdamped evolution equations
for (order-parameter) strain-tensor components, for  ferroelastic
transitions of different symmetries?

A wide variety of dynamical models have been used to date.  These
include a $2D$ or $3D$ $TDGL$ dynamics in morphological variables
for structural variants,  with a long-range potential between
squares of these variables \cite{Khatch}, motivated by ideas of
elastic inclusions \cite{eshelby}.  Other work in $2D$ and $3D$
has used a $TDGL$ equation for the $OP$ strains only, with a long-range
potential emerging between the $OP$ strains themselves
\cite{ourRC,our3DPRL,Yama} (not their squares),
by optimizing non-OP strains.  Some
authors \cite{Anantha} assumed the validity of a
$2D$ Bales-Gooding $(BG)$ form from a Lagrangian without non-$OP$
strains, and then phenomenologically added a long-range
potential between squares of the $OP$ strains.
A $TDGL$ equation in the displacements has
also been
used \cite{Onuki}. Yet other models considered displacement accelerations
equated to displacement-gradients of the free energy as forces; or
strain $TDGL$
equations coupled to vacancy field dynamics \cite{Madan}.
Finally the Lagrange-Rayleigh procedure
\cite{BG,Reid,LL} has recently
been applied \cite{JacobsHO,JacobsSR}
in $2D$, yielding an underdamped dynamics
for the displacement,
that is truncated to  overdamped equations, that are seemingly
different from the $TDGL$ form.

While these and other models yield valuable insights into
ferroelastic texturing (i.e., single-crystal microstructure), there
is clearly a need, through explicit derivation, to obtain  an
underdamped, symmetry-specific $OP$ strain dynamics for $D>1$; to
find the precise form of long-range potentials (if any) that
emerge; and to determine the regime of validity (if it exists) of
some form of $TDGL$ equations.

In this paper, we use the Lagrange-Rayleigh variational procedure
to derive a ferroelastic strain dynamics (including noise terms).
A central role is played by the $N_c$  St. Venant compatibility
conditions \cite{venant,compat,Baus,Kartha} for the $N=N_{op}+n$
symmetry adapted strains, that enforce the absence of defects (and
lattice integrity) at each instant,
and allow the $n$ non-OP strains to be expressed in terms of $N_{op}$
order parameter strains.  We show that: (i) an underdamped set of
$N_{op}$ equations can be obtained for the $OP$ strains alone, that
is of a generalized $BG$ form, with naturally emerging anisotropic
long-range (ALR) contributions to $OP$ potentials,
friction and noise. (For the $N_{op}=1$ case these are in general also
explicitly {\it retarded} in time.) (ii) The {\it same} $OP$ equations
can be
obtained, either by varying the {\it displacement}, or by varying the
{\it strains} subject
to the compatibility constraint through dynamic Lagrange multipliers.\
(iii) Dropping
strain inertial terms yields strain
$TDGL$ equations,
with nonlocal Onsager coefficients, that reduce to constants for
special friction values, resulting in a local $TDGL$ dynamics.
These act as a late-time dynamics for the damping envelope of textural
oscillations.

We explicitly demonstrate (i), (ii), (iii) above for the $2D$ triangular
to centered rectangular or $TR$ lattice transition $(N_{op}=2, n=1,
N_c=1)$ and
for the square to rectangular or $SR$ lattice transition $(N_{op}=1, n=2,
N_c=1)$, as well as present
dynamics for all other allowed $2D$ symmetries.  The procedure can be
generalized to $3D$, e.g. the cubic to tetragonal $(N_{op}=2, n=4,
N_c=6)$ transition where (static) compatibility potentials in a
$TDGL$ dynamics produce  rich textures \cite{our3DPRL}.  Our central
result is a generalized $BG$ dynamics  written in the $OP$
strains $\{\varepsilon_\ell\}$, $\ell = 1,2,...,N_{op}$ only:

$$\rho_0\ddot{\varepsilon_\ell}=\frac{{c_{\ell}}^2}{4} {\vec{\Delta}^2}
\left(
\frac{\delta
(F+F^{c})}{
{\delta\varepsilon_\ell}} +\frac{\delta\left (
R+R^c\right)}
{\delta
\dot{\varepsilon_\ell}} \right)$$
$$ + \tilde{g}_{\ell} + \tilde{g}^{c}_{\ell}, \eqno(1.3)$$
where $c_{\ell}$ is a symmetry-specific constant, and $\rho_0$ is
a scaled mass density. $F^{c}
(\{\varepsilon_\ell \})$, $R^{c}(\{\dot{\varepsilon}_\ell \})$
are the
compatibility-induced symmetry-specific contributions that emerge
naturally from the non-$OP$ free energy as additions to the $OP$ free
energy $F$ and $OP$ Rayleigh dissipation ${\it R}$, while
$\tilde{g}^{c}_{\ell}$ is the corresponding noise term that adds
to the $OP$ noise $\tilde{g}_{\ell}$.
In equation (1.3) and subsequently, we use the symbol $\vec{\Delta}$ to
denote dimensionless {\it discrete} derivatives on a reference lattice \cite{lattice}.
The generalized $BG$ equations can be written
as Langevin equation for $\varepsilon_{\ell}(\vec{k},t)$ and
$v_{\ell} \equiv \dot{\varepsilon}_{\ell}(\vec{k},t)$, yielding statistically
equivalent Fokker-Planck equations for the probability
$P(\{\varepsilon_{\ell}, v_{\ell}\}, t)$.

In the strain-variation derivation of the $BG$ dynamics above,
we introduce the concept of a strain mass-density tensor whose
components (in Fourier space) behave as $\rho_{s s'} (k) \sim \rho_0
/k^2$, that is responsible in coordinate space, for the Laplacian
on the right-hand side of (1.3).
This generalization of the $1D$ case expresses the physical idea that
long-wavelength strains
are extended lattice deformations and hence have greater inertia. We present
a physically illuminating analogy of (generalized) $BG$ dynamics as an
array of coupled nonlinear $\vec{k}$-space oscillators that have an
intrinisically hierarchical equilibration, with large
${k}$ oscillators damping out first.

The $N_{op}$ strain order-parameter equations with derived
anisotropic
long-range terms are equivalent to the $D$ displacement equations
that do not explicitly have such terms.
The advantage of the $OP$ strain approach is
that it displays and uses such  anisotropic
long-range correlations, that are
valuable in understanding simulated textures, as demonstrated below.

More generally, in the displacement $(\vec{u})$ picture, the strains are
derived quantities and the compatibility condition is an incidental
identity,
expressing the single-valuedness of $\vec{u}$.  This is analogous to describing
magnetic problems in terms of the vector potential $\vec{A}$, with $\vec{B}$
just a label for $\vec{\Delta}\times \vec{A}$, and with $\vec{\Delta}\cdot
(\vec{\Delta}\times\vec{A})=0$ just expressing a vector identity.
By contrast, in the strain-only picture, the strain tensor components
$E_{\mu \nu}$ are the physical variables, and the
compatibility conditions $\vec{\Delta}\times\left (\vec{\Delta}\times
E \right )^T =0$ (with $T$ denoting transpose) are treated as independent 
field equations expressing the
physical constraint of no defects.  This is analagous to working with
the magnetic induction $\vec{B}$, where the Maxwell's field equation
$\vec{\Delta}\cdot\vec{B} =0$ expresses the absence of magnetic monopoles.
The compatibility equation for strain has been used for a consistent 
description of forces in liquids \cite{Baus};
here we use it to develop a consistent OP dynamics for ferroelastics.

The basic idea is quite simple. Compatibility implies non-$OP$
strains $\{e_i\}$ are proportional in Fourier space to $OP$ strains
($\varepsilon_\ell$):

$$ e_i = \sum_{\ell} S_{i \ell} \varepsilon_{\ell}.
\eqno(1.4)$$

\noindent Hence, the non-$OP$ free energy $f$ (and Rayleigh dissipation),
harmonic in the non-$OP$ strains (and time derivatives),
can be written in terms of $OP$ strains. With elastic constants $a_i$,

$$ f=\frac{1}{2}  \sum_{i,\vec{k}} a_i |e_i|^2 =\frac{1}{2} \sum_{\ell, \ell' \vec{
k}} U_{\ell
\ell'}
\varepsilon_{\ell} {\varepsilon_{\ell'}}^* \equiv
F^{c} (\{\varepsilon_\ell \}).
\eqno(1.5a)$$
This defines the compatibility kernels
$$U_{\ell \ell'}=\sum_i a_i S_{i \ell} {S_{i
\ell'}}^* \eqno(1.5b)$$
for
$F^c$ (and similarly for $R^c$) of (1.3) in the  desired $OP$-only dynamics.
Thus the problem reduces to finding
the proportionality constants
$S_{i \ell}$, for each symmetry-based phase transition.

The plan of the paper, with self-contained Sections, is as
follows.  In Sec. II the $OP$ dynamics for the $TR$ case is derived by
$\vec{u}$ variation. In Sec. III we demonstrate that the same $TR$
dynamics is obtained
by strain variation, with enforced compatibility. Results for the SR case
are stated and analyzed, with derivation details in Appendix A. Numerical
simulations of some interesting $BG$ dynamic evolutions are presented
(using  standardized scaled
energies \cite{Fscaling} and dissipations,
as  in Appendix B). Noise contributions are derived, and a Fokker-Planck
formalism is set up. Section IV
discusses the equivalent inhomogeneous oscillator description of the $BG$
dynamics.
Section V deals with its $TDGL$
truncation, also derived in
Appendix C from truncated displacement dynamics
\cite{JacobsHO,JacobsSR}.
Section VI presents the compatibility kernels for other $2D$ symmetries.
Finally, Sec. VII contains a summary and discussion.

\section{ORDER PARAMETER STRAIN DYNAMICS BY DISPLACEMENT VARIATION}

Consider a Lagrangian density $\mathit{L}(\alpha ,\dot{\alpha} )=\int dt
\sum_{\vec{r}} (T-V)$ that depends on a variable $\alpha (\vec{r}, t)$
through a
kinetic  term $T= T(\dot{\alpha})=
\sum_{\vec{r}}\frac{1}{2}\rho_{0}\dot{\alpha}^2$,
and a potential term $V=V(\alpha )$.
Then, with a Rayleigh dissipation \cite{LL}
$R^{tot}=\int d\vec{r}\frac{1}{2} {\eta}
\dot{\alpha}^2$ (where $\eta$ is the friction coefficient), we have, by
variation in $\alpha$, the Lagrange-Rayleigh equation:

$$\frac{d}{dt}\frac{\partial{\mathit{L}}}{\partial\dot{\alpha}}-
\frac{\partial{\mathit{L}}}{\partial\alpha}=-
\frac{\partial R^{tot}}{\partial\dot{\alpha}}. \eqno(2.1)$$

In this Section we work in the displacement picture, and
consider variations in displacement
$\alpha (\vec{r},t) \longrightarrow { u_{\mu}(\vec{r},t) }
 $
that will generate  $\mu = 1,..,D$ equations, for $\ddot{u}_\mu$
in $D$ dimensions.
The potential or
Gibbs
free energy $V$ depends on displacement derivatives, that are $N=N_{op}+n$
symmetry-adapted linear combinations of the symmetric
Cauchy strain tensor

$$E_{\mu \nu}=\frac{1}{2}\left( \Delta_{\mu} u_{\nu} +\Delta_{\nu}
u_\mu\right). \eqno(2.2)$$

\noindent We neglect `geometric' nonlinearity, as justified by the scaling
of  Appendix B, where such corrections are  higher order in a typical
(small) strain value.
We notationally distinguish between $OP$ strains $\{\varepsilon_\ell\}$
and non-$OP$ strains $\{ e_i\}$.  The potential
$V=F(\{\varepsilon_\ell\})+f(\{e_i\} )$ is anharmonic through
$F$
in the $N_{op}$ order parameter strains, and harmonic through $f$ in the
$n$ non-$OP$ strains,  while the Rayleigh dissipation
$\mathit{R^{tot}}(\{\dot{\varepsilon_\ell}\},\{\dot{e_i}\})$
is harmonic
in both the strain rates:

$$f =
\sum_{\vec{r}, i}
\frac{1}{2}a_i e_i^{2}   ; \ \
R^{tot}=\sum_{\vec{r},\ell}\frac{1}{2}A'_{\ell}
\dot{\varepsilon}^2_\ell +\sum_{\vec{r},i}\frac{1}{2}a'_i\dot{e_i}^2.
\eqno(2.3)$$

\noindent Here as in Appendix B, $\{A_\ell\}$,
$\{a_i\}$ and  $\{A'_\ell\}$, $\{a'_i\}$ are,
respectively,  $OP$ and non-$OP$ second order elastic  and friction
coefficients, and the sum
is over sites $\{\vec{r}\}$ of a reference lattice, while $t$ is a scaled 
time. \\

\noindent{\it TR dynamics from displacement variation}

Consider the  triangular to centred rectangular lattice
($TR$) transition, for
which $N_{op}=2$ and $n=1$. Figure 1 shows the $TR$,  the square to
rectangular or $SR$, and other lattice transitions. (While it is true
\cite{JacobsHO,JacobsSR} that these
correspond to $2D$ projections/analogs of hexagonal to orthorhombic, and
tetragonal to orthorhombic lattice transitions, respectively, we
will reserve this $3D$ terminology for full $3D$ analyses, elsewhere.)
The symmetry-adapted non-$OP$ compressional strain 
$e_1=\frac{1}{2}\left(\Delta_x u_x+\Delta_y u_y\right)$, whereas the 
$OP$ are the `deviatoric'
$\varepsilon_{2}=\frac{1}{2}\left(\Delta_x u_x-\Delta_y u_y\right)$ and
shear strain
$\varepsilon_{3}=\frac{1}{2}\left(\Delta_x u_y+\Delta_y u_x\right)$. 
The deviatoric strain can be regarded as a diagonal shear. 

Then 
(2.3) for the non-OP compressional energy is 
$f=\sum_{\vec{r}}\frac{1}{2}a_{1}{e_1}^2$,
$R^{tot}=\frac{1}{2}\sum_{\vec{r}} [ a'_{1}{\dot{e}_1}^2
+( A'_2{\dot{\varepsilon}_2}^2
+ A'_3{\dot{\varepsilon}_3}^2)].
$
The anharmonic and fourth-order (triple-well) free energy $F$ for 
$\varepsilon_2$, $\varepsilon_3$, is given  below in  (3.27a),
although this explicit form is not needed in the derivation.

Defining $OP$ free energy derivatives (i.e., stresses)
$F_{2,3} \equiv \frac{\delta F}{\delta\varepsilon_{2,3}(r,t)},
$
the Lagrange-Rayleigh variation with respect to displacements $\vec{u}
(\vec{r},t)$
gives for the dynamics of (2.1)

$$ \rho_0\ddot{u}_x=  \frac{1}
{2} [a_1 \Delta_x e_1+ \Delta_xF_2 + \Delta_y F_3 ]$$
$$ +\frac{1}{2}[a'_1\Delta_x\dot{e}_1+ A'_2\Delta_x\dot{\varepsilon}_2
+A'_3\Delta_y\dot{\varepsilon}_3]; \eqno(2.4a) $$ $$
\rho_0\ddot{u}_y= \frac{1}{2} [a_1\Delta_y e_1-\Delta_y F_2+
\Delta_xF_3 ] $$
$$+\frac{1}{2} [a'_1\Delta_y\dot{e}_1- A'_2\Delta_y\dot{\varepsilon}_2+
A'_3\Delta_x\dot{\varepsilon}_3].\eqno(2.4b) $$
These displacement equations have been obtained previously
\cite{JacobsHO},
but were then truncated
by dropping the displacement acceleration (second time derivative) to
yield reduced  equations that are analyzed further in
Appendix C. Instead, we pursue here the underdamped $OP$-strain equations
and find they have a generalized $BG$ form.

The strains obey the compatibility constraints, that in the
displacement
picture ensure  that $\vec{u}$ is single-valued (i.e., cross-derivatives
commute).  The equation
for the $2D$ case is, at every instant,
$$\vec{\Delta}^2e_1 -\left(\Delta_{x}^{2}-
\Delta_{y}^2 \right) \varepsilon_2
- 2\Delta_x\Delta_y\varepsilon_{3}=0.\eqno(2.5)$$

Taking spatial derivatives of (2.4) we obtain the full 
underdamped
equations for the
strains:
$$\rho_0\ddot{e_{1}} = \frac{1}{4}[ a_1\vec{\Delta}^2e_{1}
+({\Delta_x}^2-{\Delta_y}^2)\frac{\partial F}{\partial \epsilon_{2}} +
2\Delta_{x}\Delta_{y}\frac{\partial F}{\partial
\epsilon_{3}}]$$
$$+ \frac{1}{4}[a_{1}'\vec{\Delta}^2\dot{e}_{1} +
A_{2}'({\Delta_x}^2-{\Delta_y}^2)\dot{\varepsilon}_{2}
+2A_{3}'\Delta_{x}\Delta_{y}\dot{\varepsilon}_{3}];
\eqno(2.6a)$$
$$\rho_0\ddot{\varepsilon_{2}} =
\frac{1}{4}[a_1 ({\Delta_x}^2-{\Delta_y}^2)e_{1}
+\vec{\Delta}^2\frac{\partial F}{\partial \varepsilon_{2}}]$$
$$+\frac{1}{4}[a_{1}'({\Delta_x}^2-{\Delta_y}^2)\dot{e}_{1}
+A_{2}'\vec{\Delta}^2\dot{\varepsilon}_{2}];
\eqno(2.6b)$$
$$\rho_0\ddot{\varepsilon_{3}} =
\frac{1}{4}[2 a_{1}\Delta_{x}\Delta_{y}e_{1}
+
\vec{\Delta}^2\frac{\partial F}{\partial \varepsilon_{3}}]$$
$$+ \frac{1}{4} [2 a_{1}'\Delta_{x}\Delta_{y}\dot{e}_{1}
+ A_{3}'\vec{\Delta}^2\dot{\varepsilon}_{3}].
\eqno(2.6c)$$

By taking appropriate derivatives of (2.6), it is easy to see that
the compatibility condition (2.5) is satisfied as an identity. This
{\it linear}  equation, (2.5), can then be used instead of the
nonlinear (2.6a), to eliminate the non-$OP$ strain
$e_1(\vec{k},\omega )$ in terms of the $OP$ strains
$\varepsilon_{2,3}(\vec{k},\omega )$, assuming periodic boundary
conditions.

Defining $a_{1\omega}\equiv a_{1} - i\omega a_{1}'$,
transforming to Fourier space, and with $F_{2,3}$
now defined as
$F_{2,3} \equiv \frac{\partial F}{\partial
\varepsilon_{2,3}^{\star}(\vec{k},
\omega)},$
the compatibility constraint (2.5)

$$ Q_1 e_{1}(\vec{k},\omega) + Q_2 \varepsilon_{2}(\vec{k},\omega)
+ Q_3 \varepsilon_{3}(\vec{k},\omega) = 0 \eqno(2.7)$$

with
$Q_1 \equiv -k^2, Q_2 \equiv {k_x}^2 -{k_y}^2,
Q_3 \equiv 2 k_x k_y$,
is used to eliminate $e_{1}(\vec{k},\omega)$, giving
$N_{OP}=2$ order parameter strain-only equations \cite{lattice}: 
$$\rho_0\omega^{2}\varepsilon_{2} = \frac{k^2}{4}\left[ a_{1\omega}
\left(\frac{Q_{2}^2}{Q_{1}^2}\varepsilon_{2} + \frac{Q_{2}Q_{3}}{Q_{1}^2}
\varepsilon_{3}\right) +
F_{2} - {i\omega}A_{2}'\varepsilon_{2} \right],
\eqno(2.8a)
$$
$$\rho_0\omega^{2}\varepsilon_{3} = \frac{k^2}{4} \left[ a_{1\omega}
\left(\frac{Q_{2}Q_{3}}{Q_{1}^2}\varepsilon_{2} + \frac{Q_{3}^2}{Q_{1}^2}
\varepsilon_{3} \right) +
F_{3} - {i\omega}A_{3}'\varepsilon_{3}\right]. \eqno(2.8b)
$$

These can be succinctly written as an $OP$ dynamics
for $\varepsilon_{\ell}(r,t)$
with $(\ell =2,3)$:
$$\rho_0 \ddot{\varepsilon}_\ell=\frac{1}{4}{\vec{\Delta}^2} \left(
{{\delta
(F+F^{c})}\over
{\delta\varepsilon_\ell}}+{{\delta\left ( \mathit{R}+\mathit{R^c}\right)}
\over{\delta
\dot{\varepsilon}_\ell}} \right), \eqno(2.9)$$
where $F^c$ and $R^c$ are compatibility-induced
contributions from the non-$OP$ free energy $f$ and dissipation $R$
written in terms of the $OP$
strains:

$$
F^{c} =
\frac{1}{2}a_{1}\sum_{\vec{k},\ell,\ell'}{\it
U}^{c}_{\ell \ell'}(\hat{k})\varepsilon_{\ell}(\vec{k},t)
{\varepsilon^*}_{\ell'}(\vec{k},t), \eqno(2.10a)$$

$${R}^{c} = \frac{1}{2}a_{1}' \sum_{\vec{k},\ell,\ell'}
{\it
\eta}^{c}_{\ell \ell'}(\hat{k})\dot{\varepsilon}_{\ell}(\vec{k},t)
\dot{{\varepsilon^*}}_{\ell'}(\vec{k},t),
\eqno(2.10b) $$
where the orientation-dependent kernel \cite{lattice}
${\it U}^{c}_{\ell \ell'}(\hat{k})={\it \eta}^{c}_{\ell \ell'}(\hat{k})$ 
is defined implicitly
above, and given explicitly in (3.29) of Sec. III,  following a
strain-based derivation of the same dynamics.

The system of $N_{op}=2$ underdamped equations (2.9) derived for the
strains is clearly of a
Bales-Gooding form [compare with (1.1)], but now
generalized in three ways:  by the derived replacement
$\frac{\partial^2}{\partial {x}^2}\rightarrow \vec{\Delta}^2$;
by the appearance of a
compatibility-induced anisotropic long-range ($ALR$) interaction, between
$\varepsilon_\ell$-${\varepsilon}_{\ell'}$
(and not $\varepsilon^2_{\ell}$-$\varepsilon^2_{\ell'}$);
and of a similiar compatibility-induced $ALR$ dissipation,
between ${\dot{\varepsilon}}_{\ell}$-${\dot{\varepsilon}}_{\ell'}$
strain rates.

An interesting consequence of the $BG$ structure with periodic
boundary conditions is that the
``$\vec{k}=0$'' (or more precisely, $\vec{k}\rightarrow 0$)
$OP$ strain obeys $\rho_0 \ddot{\varepsilon_{\ell}}(\vec{k}=0,t)=0$,
so macroscopic strain momentum 
$\rho_0 \dot{\varepsilon_{\ell}}(\vec{k}=0,t)=0$ is conserved
\cite{bc}, unaffected by internal forces and dissipations.
The solution is $\varepsilon_{\ell}(\vec{k}=0,t)  = 
\dot{\varepsilon_{\ell}}(\vec{k}=0,0) t +
\varepsilon_{\ell}(\vec{k}=0,0)$. In the special case of an austenite
phase with initial
conditions $\dot{\varepsilon_{\ell}}(\vec{k}=0,0)=0$, and
$\varepsilon_{\ell}(\vec{k}=0,0) \sim \sum_{r}\varepsilon_{\ell}(r,0)=0$,
strains of $\it{both}$ signs will develop on cooling below
transition, as a consequence of the dynamics. The $OP$ strain is in this 
case like a `charge' that is generated in sign-balancing pairs, and 
the notion of elastic
`screening' helps in understanding simulations presented later.

The square-rectangle ($SR$) dynamics, driven by a deviatoric strain
$OP$, can similiarly be shown by displacement variation to be also of
the generalized $BG$ form, as given in the first part of
Appendix A.
However, we now proceed to derive the strain dynamics through strain
variation.

\section {OP STRAIN DYNAMICS BY STRAIN VARIATION WITH COMPATIBILITY
CONSTRAINTS}

In discussions of ferroelastics, it is common to assert that although the
free energy is in terms of the strains,
the true basic variables for such systems are displacements,
since strains are just displacement derivatives.
Thus Monte Carlo simulations, numerical solutions of dynamic equations and
static analyses of textures, even when expressed in terms of strains,
are all finally  performed in terms of
displacements.
Following the electromagnetic analogy
mentioned in the Introduction, an alternative treatment  is
in terms of {\it strains} as the basic variables.
The free energy for a first order transition, say, $F_0 \sim \varepsilon^6$,
is then regarded as zeroth order in
derivatives, whereas in the displacement picture it is a sixth power of
derivatives. In this Section, we
derive ferroelastic dynamics for the $TR$ and $SR$
cases, using  strains as the variational quantities.  Results for external
stress and noise are also stated.
We use periodic boundary conditions throughout \cite{bc,surface} and
transform 
between
coordinate $\vec{r}$ and wavevector $\vec{k}$ descriptions, as
convenient. The Lagrange-Rayleigh dynamics equations for a general variable
$\alpha$ is

$$\frac{d}{dt}\frac{\partial{\mathit{L}}}{\partial\dot{\alpha}}-
\frac{\partial{\mathit{L}}}{\partial\alpha}=-
\frac{\partial{R^{tot}}}{\partial\dot{\alpha}},
\eqno(3.1)$$
and we consider here the strains as the variables:

$$\alpha\rightarrow \{ E_{\mu\nu}(\vec{r},t)\}. \eqno(3.2)$$

\begin{figure}[h]
\epsfysize=9cm
\epsffile{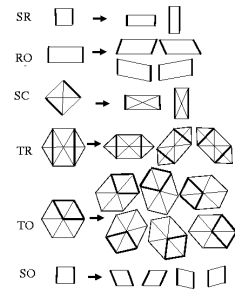}
\caption{Symmetry-allowed transitions in  $2D$ for the four crystal systems.
The dark lines are guides to the eye, for deformations. There is a
one-component strain order parameter for: (a) the square to
rectangle or $SR$ case, driven by $\varepsilon_2$; (b) the rectangle
to oblique or $RO$ case driven by $\varepsilon_2$; and (c) the square
to centered rectangle or $SC$ case, driven by $\varepsilon_3$.
A two-component $OP$, or {\it two} one-component $OP$'s, lead(s) to
(d) the triangular to centered rectangle or $TR$
case, driven by $\varepsilon_2, \varepsilon_3$; (e) the triangle to
oblique or $TO$ case, driven by
$\varepsilon_2, \varepsilon_3$; and
(f) the square to oblique or $SO$ case, driven by $\varepsilon_2$ and
$\varepsilon_3$, independently. [The oblique to oblique ferroelastic
($P2$ to $P1$) transition that involves merely
a loss of inversion symmetry is not considered \cite{Hatch}.]}
\label{fig1}
\end{figure}

 In total, if we consider all space groups in $2D$,
there are 23 ferroelastic transitions \cite{Hatch}
in 2D. Figure 1 shows the six symmetry-allowed transitions in $2D$ for the four
crystal systems.  However, of these only six correspond to mono-atomic basis.
We will present the initial part of the argument in general form,
before focusing on the triangular to (centered) rectangular lattice  or $TR$
case; and the square to rectangle or $SR$ case, with other $2D$ transitions
considered in Sec. VI.

The Lagrangian contains the potential $V$ and kinetic energy $T$ that
depend on symmetry-adapted strains or linear combinations of the strain
tensor $\{E_{\mu\nu}\}$, that are irreducible representations of the
unit-cell symmetry group. The $s=1,2,\ldots,N$ symmetry-adapted strains 
are written as $\{e_s\}$, with $N=3$ in $2D$ (and 6 in $3D$).  In 2D the 
compressional $(e_1)$, deviatoric $(e_2)$ and shear $(e_3)$ strains are 
defined by
$$\frac{e_1}{c_1}={1\over 2}(E_{xx}+E_{yy}), \  \ \ \
\frac{e_2}{c_2}={1\over
2}(E_{xx}-E_{yy});$$
$$\frac{e_3}{c_3}=\frac{1}{2}(E_{xy}+E_{yx}), \eqno(3.3)$$
where $c_1,c_2,c_3$ are symmetry-specific constants.

The strain tensor
(neglecting geometric nonlinearity as justified in Appendix B)
obeys the St. Venant compatibility condition, that is here a field
equation  forbidding defects
such as dislocations and vacancies \cite{compat,Baus} at every instant:
$$\vec{\Delta}\times\left(\vec{\Delta}\times {\b{\b{E}}} (\vec{r},t)
\right)^T = 0,
\eqno(3.4a)$$
$$\vec{k}\times {\b{\b{E}}} (\vec{k},t) \times \vec{k} =0,
\eqno(3.4b)$$
with no source term on the right hand side. The $p=1,2,\ldots, N_c$
compatibility conditions are:
$$C^{(p)}\equiv\sum_s{\hat{Q}}^{(p)}_s e_s ({\vec{r}},t )=0,
\eqno(3.5)$$
where ${\hat{Q}}^{(p)}_\mu$ are second-order derivative operators,
from (3.4a).
In $2D$ there is only one compatibility equation $N_c=1$, and
the operators in
Fourier space are $Q_1 (\vec{k})\equiv - {\vec{k}^2}/{c_1}$,
$Q_2(\vec{k}) \equiv {({k_x}^2 - {k_y}^2)}/{c_2}$,
$Q_3(\vec{k}) \equiv  {2 {k_x} {k_y}}/{c_3}$. Thus
(3.5)  for $e_{s} (\vec{k},t)$ is, from (3.4b),
$$Q_1 e_1 + Q_2 e_2  +Q_3 e_3=0.
\eqno(3.6)$$

\noindent The anisotropic compatibility factors $Q_{1,2,3} (\vec{k})$
encode the discrete symmetries of the compression, deviatoric and
shear strains. The
symmetry constants for the $TR$ case are $c_1 = c_2 = c_3
=1$, whereas for the $SR$ case they are $c_1= c_2 = \sqrt{2}, \ c_3 =1$.
The compatibility constraint will be invoked repeatedly in the
derivations below. The physical meaning of the constraint is that order
parameter strains should not tear, or cause defects in the
lattice, i.e. lattice integrity is maintained. Suppose, in a
sea of square unit cells, one cell was made rectangular (local deviatoric
strain). It is clear that in order to maintain lattice integrity,
the neighboring cells must also deform, inducing
all three strains in an inter-related way, with 
 a similar (but smaller) deformation of
the larger number of further neighbors. The requirement
of a smooth, compatible fitting together of neighboring unit cells
will cause the disturbance to propagate outwards, and in an anisotropic way
(due to discrete crystal symmetry and elastic constants): the {\it local}
condition has {\it global} consequences.

The Lagrangian $L$ also contains the compatibility constraints through
dynamic Lagrange
multipliers $\{ \Lambda^{(p)}({\vec{r}},t )\}$:
$$L=\int dt\left[
T-V-\sum_{p,{\vec{r}}}\Lambda^{(p)}C^{(p)}\right]. \eqno(3.7)$$

We now: (a) obtain the kinetic energy $T (\{{\dot{e}}_{\mu}\})$ in terms of
the
time derivatives
of the symmetry-adapted strains; and (b) use this to derive the $TR$ and
$SR$
dynamics by strain variation, incorporating the compatibility
constraint.

\subsection{ Kinetic energy in terms of strain rates:}

Since Newtonian dynamics is for {\it point} particles, the kinetic energy
in terms of displacements is:
$$T={1\over 2}\rho_0\sum_{{\vec{r}},\mu} {\dot{u}}_\mu^2({\vec{r}},t
) ={1\over 2}\rho_0\sum_{{\vec{k}},\mu} |\dot{u}_\mu ({\vec{k}},t) |^2
\eqno(3.8)$$
in coordinate and wave-vector spaces,
where $\rho_0$ is a dimensionless mass-density that
is a ratio of  typical kinetic and elastic
energy densities (Appendix B).
In the strain picture, the displacement can be derived from the strains 
through the Kirchoff-Cesaro-Volterra relation \cite{Baus}
$$\vec{u}(\vec{r})= \int_{C(\vec{r}_o,\vec{r})}
[{\bf E}(\vec{l}) +
\{(\vec{l}-\vec{r}) \times
\vec{\nabla}_{\vec{l}}\times {\bf E} (\vec{l})\}].d{\vec{l}},
\eqno(3.9a)$$
where the line integral is along any contour $C(\vec{r}_o,\vec{r})$ to
$\vec{r}$ from $\vec{r}_o$, that is a fixed point of the deformation.
Taking derivatives with respect to $\vec{r}$, the
symmetric
combination of derivatives is

$$ \frac{1}{2}( \Delta_{\mu}u_{\nu} + \Delta_{\nu}u_{\mu}) =
E_{\mu\nu}.
\eqno(3.9b)$$

In $2D$, using (3.3) this gives
$$\Delta_xu_x=\frac{e_1}{c_1} +\frac{e_2}{c_2};
\ \ \Delta_yu_y=\frac{e_1}{c_1}-\frac{e_2}{c_2}.
\eqno(3.10a)$$
Taking time derivatives and transforming to Fourier space,
$${\dot{u}_x}(\vec{k},t)=\frac{ \frac {\dot{e}_1} {c_1}
+\frac {\dot{e}_2} {c_2}  }
{ik_x};
\ \
{\dot{u}_y}(\vec{k},t)=\frac{\frac{\dot{e}_1}{c_1}-\frac{\dot{e}_2}
{c_2}}{ik_y}.\eqno(3.10b)$$
Inserting (3.10b) into (3.8) yields the kinetic energy,  that is
{\it nonlocal}
in terms of
the strain rates:

$$
T = \sum_{\vec{k},s,s'}\frac{1}{2}\rho_{s s'}(\vec{k})
\dot{e_{s}^{\star}}(\vec{k} ,t)
\dot{e}_{s'}(\vec{k},t)    $$

$$=
\sum_{\vec{r},\vec{r'}s,s'}\frac{1}{2}\rho_{s s'}(\vec{r}-\vec{r'})
\dot{e_{s}^{\star}}(\vec{r} ,t)
\dot{e}_{s'}(\vec{r'},t),
\eqno(3.11) $$
where we have introduced an anisotropic `strain mass-density tensor'
whose components turn out to be related to ratios of the compatibility
factors of (3.6):

\[
\rho_{s s'}(\vec{k})
=\rho(k)
\left[
\begin{array}{cc}
\frac{1}{c_{1}^{2}}\frac{k^2}{k_{x}^{2} k_{y}^2}                     &
-\frac{1}{c_{1}c_{2}}\frac{(k_{x}^2 - k_{y}^2)} {k_{x}^{2} k_{y}^2}
\\
-\frac{1}{c_{1}c_{2}}\frac{(k_{x}^2 - k_{y}^2)}{k_{x}^{2} k_{y }^2}  &
\frac{1}{c_{2}^2}\frac{k^2}{k_{x}^{2} k_{y}^2}
\end{array}\right]
\]

\[
\ \ \ ={\rho(k)}
\left[
\begin{array}{cc}
        (\frac{Q_1}{Q_3})^2                     &
             (\frac{Q_1 Q_2}{{Q_3}^2})        \\
          (\frac{Q_2 Q_1}{{Q_3}^2})  &
\frac{c_{1}^2 }{c_2^2}  (\frac{Q_1}{Q_3})^2
\end{array}\right].
\ \ \ \ \ \ \ \ \ \ \ \ \ \ \ \ \ \ \ \ \ \ \ \ \ (3.12a)
\]
This strain mass density
tensor is a kinematic time-independent quantity
true for all $2D$ symmetries
and has a determinant $(2/c_{1}c_{2}k_{x}k_{y})^2$.
Here
$$\rho (k) \equiv  \frac{4\rho_0}{c_{3}^2k^2} \eqno(3.12b)$$
and therefore
long-wavelength strains
over many lattice spacings are effectively more `massive', as is physically
reasonable. In  $2D$ coordinate space, $\rho_{s s'}({\vec{R}})\sim (4
\rho_0) ln (|\vec{R}|)$. It is this inverse Laplacian
dependence that gives rise
to the Bales-Gooding structure (1.3) of the underdamped dynamics.
The strain kinetic energy can be expressed only in terms of the
compressional and
deviatoric strain rates. i.e., the `shear components' of the strain mass
tensor are zero, $\rho_{s 3} = \rho_{3 s} =0$. This is because
from (3.3) and (3.9b), the shear strain rate
$$ \frac{{\dot{e}}_3({\vec{k}})}{c_3}  =
{1\over 2}[ik_x {\dot{u}}_y({\vec{k}})+ik_y{\dot{u}}_x({\vec{k}})],
\eqno(3.13)$$
is not independent, but is related to the  other strain rates by
a consistency condition through (3.10b), that turns out to be precisely
the compatibility constraint.

\subsection{Dynamics by strain variations:}

It is convenient to henceforth notationally distinguish between $n$
non-order
parameter strains $\{e_i\}$ and $N_{op}$ order parameter
strains $\{\varepsilon_\ell\}$.
The compatibility conditions (3.5) become
$$C^{(p)}{(r,t)}=\sum_\ell
\hat{Q}_{\ell}^{(p)}\varepsilon_\ell({\vec{r}},t)
+\sum_{i}{\hat{Q}}^{(p)}_i e_i(\vec{r},t)=0, \eqno(3.14a)$$
and therefore in Fourier space
$$C^{(p)}({\vec{k}},t )=\sum_{\ell}Q^{(p)}_\ell
({\vec{k}})\varepsilon_\ell
({\vec{k}},t )+\sum_{i}Q^{(p)}_i({\vec{k}})e_i({\vec{k}},t )=0. 
\eqno(3.14b)$$

The Gibbs free energy $V(\{\varepsilon_{\ell}\}$,$\{e_i\})$ depends
harmonically on the non-$OP$ strains through $f$ and anharmonically on
$OP$ strains through $F$, whereas the Rayleigh dissipation $R^{tot}$ to
the
lowest order depends harmonically on all strain rates.
Thus,

$$V=f(\{e_i\})+F(\{\varepsilon_\ell\}); ~~~f={1\over 2}\sum_{i, \vec{r}}
a_ie^2_i;$$
$$R^{tot}={1\over 2}\sum_{i, \vec{r}}{a_i}'\dot{e_i}^2+{1\over
2}\sum_{\ell,\vec{r}} A_{\ell}'
{\dot{\varepsilon}}_{\ell}^{2},  \eqno(3.15)$$
where $a_i,a'_i$ are the non-$OP$ elastic and friction constants.

The kinetic energy in terms of strain rates  $\{\dot{e_i}\},\{
\dot{\varepsilon}_\ell\}$, from (3.11), is
$$
T =  \sum_{\vec{k},\ell,\ell'}\frac{1}{2}\rho_{\ell \ell'}
(\vec{k})
\dot{\varepsilon}_{\ell}^{\star}(\vec{k} ,t)
\dot{\varepsilon}_{\ell'}(\vec{k},t)
$$
$$ +
\sum_{\vec{k},i,i'}\frac{1}{2}\rho_{ii'}(\vec{k})
\dot{e_{i}^{\star}}(\vec{k} ,t)
\dot{e_{i'}}(\vec{k},t)   \}.  \eqno(3.16) $$
Using (3.14b), (3.15) and (3.16), the
Lagrange-Rayleigh dynamics for $OP$ and non-$OP$
strains are given by directly varying  the {\it strains} in (3.1) and
(3.7):
$$
\sum_{\ell'}\rho_{\ell \ell'}\ddot{\varepsilon}_{\ell'} +
\sum_{i'}\rho_{\ell i'}\ddot {e}_{i'}
= -\frac{\partial F}{\partial\varepsilon_{\ell}} -
\sum_{p}Q_{\ell}^{(p)}\Lambda^{(p)}
- A_{\ell}'\dot{\varepsilon_{\ell}}; \eqno(3.17a)$$
$$
\sum_{i'}\rho_{ii'}\ddot{e_{i'}} + a_{i}e_{i} + a_{i}'\dot{e_{i}} +
\sum_{p} Q_{i}^{(p)}\Lambda^{(p)}
=-\sum_{\ell'}\rho_{i\ell'}\ddot{\varepsilon}_{\ell'};
\eqno(3.17b)$$
$$
\sum_{i}Q_{i}^{(p)}e_{i} = -\sum_{\ell}Q_{\ell}^{(p)}\varepsilon_{\ell}
. \eqno(3.17c)$$
We have written the equations in a general form for future
$3D$ $BG$ generalizations such as the cubic to tetragonal transition
\cite{our3DPRL}. There are $(n+N_c)$ linear
equations (3.17b,c) for $(n+N_c)$ variables $\{e_i\}$,
$\{\Lambda^{(p )}\}$
that can be written in matrix form,
and inverted to yield the dynamics for the $OP$ strains
$\{\varepsilon_{\ell}\}$.
We do not pursue this general treatment here, but now specialize
to the $TR$ case
and give the result for $SR$ at the end of the section. Other symmetries
are considered in Sec. VI. \\

\noindent{\it TR underdamped dynamics:}

For the $TR$ transition, $n=1$, $N_{op}=2$, $N_{c}=1$;
the non-$OP$ strain is compressional $e_1$; the $OP$ strains are
`deviatoric' and shear ($\varepsilon_{2},\varepsilon_{3}$ respectively);
whereas the symmetry constants of (3.6) are $c_1=c_2=c_3=1$.
The strain mass-density components are $\rho_{11}=\rho (k)
(Q_1/Q_3)^2 = \rho_{22}$, $\rho_{12}= \rho (k) (Q_1 Q_2/{Q_3}^2)
=\rho_{21}$.
We have $N_{op}+n +N_c= 2+1+1=4$ equations like (3.17):
$$
\rho_{22}\ddot{\varepsilon}_{2} + \rho_{21}\ddot{e}_{1} =
-\frac{\partial F}{\partial \varepsilon_{2}^{\star}}
- Q_{2}\Lambda -A_{2}'\dot{\varepsilon}_{2}; \eqno(3.18a)$$
$$ 0 = -\frac{\partial F}{\partial \varepsilon_{3}^{\star}}
- Q_{3}\Lambda -A_{3}'\dot{\varepsilon}_{3}; \eqno(3.18b)$$
$$\rho_{11}\ddot{e}_{1}+\rho_{12}\ddot{\varepsilon}_{2}  =
-a_{1}e_{1} - a_{1}'\dot{e}_{1}-Q_{1}\Lambda;
\eqno(3.18c)$$
$$Q_{1}e_{1} = -Q_{2}\varepsilon_{2}-Q_{3}\varepsilon_{3}.
\eqno(3.18d)$$

There are $N_{op}=2$ nonlinear equations for $\varepsilon_{2,3}$
and $n+N_{c}=2$ linear equations for $e_{1}$, $\Lambda$. We can use 
the compatibility equation (3.18d) for the
strain mass density $\{\rho_{s s'}({\vec{k}} )\}$
components in (3.12) to write the expression on the LHS of (3.18c) as
$$\rho_{11}\ddot{e}_1+\rho_{12}\ddot{\varepsilon}_{2}=
\frac{4\rho_0}{Q_{3}} \ddot{\varepsilon}_3.
\eqno(3.19)$$

The Lagrange multiplier of (3.18c) is determined by the $OP$
strains as

$$\Lambda (\vec{k},\omega ) = \frac{1}{Q_{1}^2} \left[  {a_{1\omega}
\varepsilon_2}
+ \left(
a_{1\omega}{Q_3}+\frac{4\rho_0\omega^2
Q_1}{Q_3}\right)\varepsilon_3\right], \eqno(3.20)$$
where $a_{1\omega} = a_{1} - i\omega_{1}'$.
Substituting (3.20) into (3.18)
with
the identity
$$\rho_{22}-\rho_{21}\frac{Q_2}
{Q_1} = \frac{4\rho_0}{k^2},
\eqno(3.21)$$

\noindent yields the $TR$ equations in terms of the $OP$ strains alone \cite{lattice},

$$\rho_0\omega^2\varepsilon_2=\frac{k^2}{4} \left[ \frac{\delta F}{\delta
\varepsilon_2^*}+a_{1\omega} \left(
\frac{Q_2^2}{Q_1^2}\varepsilon_2+
\frac{Q_2 Q_3}{Q_1^2}\varepsilon_3 \right) -i\omega A_{2}'
\varepsilon_2 \right], \eqno(3.22a)$$
$$\rho_0\omega^2\varepsilon_{3}={\frac{k^2}{4}}\left[ \frac{\delta F}
{\delta\varepsilon_3^*}+a_{1\omega}\left(\frac{Q_2 Q_3}{Q_1^2}
\varepsilon_2+\frac{Q_3^2}{Q_1^2}\varepsilon_3 \right)
-i\omega A_{3}'\varepsilon_{3} \right]. \eqno(3.22b)$$

Since ratios of compatibility factors recur, it is useful to define
$$Q_{\ell ,\ell'}(\hat{k}) \equiv
\frac{Q_{\ell}(\vec{k})}{Q_{\ell'}(\vec{k})}. \eqno(3.23)$$

\noindent The non-OP strain in (3.18d) is a {\it derived} quantity
$$e_1({\vec{k}},\omega
)=
-\left[Q_{2,1}(\hat{k})\varepsilon_2({\vec{k}},
\omega )+
Q_{3,1}(\hat{k})\varepsilon_3({\vec{k}},\omega
)\right]
$$
$$\equiv S_{12}\varepsilon_2 + S_{13} \varepsilon_3,
\eqno(3.24)$$
defining for the $TR$ case,
the constants $S_{i \ell}(\hat{k})$ mentioned in (1.4), that
depend only on the wavevector direction $\hat{k}$, i.e., no non-trivial 
length scale \cite{lattice}, and not on its magnitude
$|\vec{k}|$ (i.e., no length scale).

In a compact form with $\ell = 2,3$,

$$\rho_0\omega^2\varepsilon_{\ell}=\frac{k^2}{4} \left (
\frac{\delta (F +F_c)}{\delta \varepsilon_{\ell}^* (\vec{k},\omega)}
+ \frac{\delta\left ( R+R^c\right)}
{\delta {\dot{\varepsilon}_\ell}^* (\vec{k},\omega)}
\right ). \eqno(3.25) $$
This is written as an $OP$ strain-only dynamics for
$\varepsilon_{\ell}(\vec{r},t)$

$$\rho_0 \ddot{\varepsilon_\ell} (\vec{r},t)=\frac{1}{4}{\vec{\Delta}^2}
\left(
\frac{\delta (F+F^{c})}
{\delta\varepsilon_{\ell}(\vec{r},t)} +
\frac{\delta\left ( \mathit{R}+\mathit{R^c}\right)}
{\dot{\varepsilon_\ell} (\vec{r},t)} \right),  \eqno(3.26)$$

\noindent that can be written as a strain-momentum continuity
equation \cite{bc}. The  $OP$ free energy from Appendix B is

$$F_0 =
\sum_{\vec{r}} {[  \frac{K_0}{2}\sum_\ell
({\vec{\Delta}}\varepsilon_{\ell})^2  +(\tau -1)}({\varepsilon_2}^2+
{\varepsilon_3}^2)$$
$$+ \{({\varepsilon_2}^2+
{\varepsilon_3}^2)
-2(\varepsilon_{2}^3 -3
\varepsilon_{2}
\varepsilon_{3}^2)+ (\varepsilon_{2}^{2}
+\varepsilon_{3}^2)^2 \} ].
\eqno(3.27a)$$
This free energy is
invariant under the 6mm point group operations. Specifically, under three 
fold
rotations,  $E_{x'x'}=\frac{E_{xx}}{4}+\frac{3E_{yy}}{4}+
\frac{\sqrt{3}}{2}E_{xy}; E_{y'y'}=\frac{3}{4}E_{xx} + \frac{1}{4}E_{yy} -
\frac{\sqrt{3}}{2}E_{xy};
E_{x'y'} =-\frac{\sqrt{3}}{4}(E_{xx}-E_{yy})-\frac{1}{2}E_{xy}$
and the non-$OP$ and $OP$ strains transform as
$e'_1=e_1, \varepsilon'_2 =-\frac{1}{2}\varepsilon_2 - \frac{\sqrt{3}}{2}
\varepsilon_3, \varepsilon'_3=\frac{\sqrt{3}}{2}\varepsilon_2
-\frac{1}{2}\varepsilon_3$.
The (anharmonic) $OP$ free energy has not been explicitly used in the
derivation of the dynamics, whose structure depends only on the
number and nature of the (harmonic) non-$OP$
strains.

The compatibility induced $OP$-$OP$ interaction
is the non-$OP$ free energy written in terms of the $OP$ strains,
$f(\{e_1(\{\varepsilon_\ell\})\}) \equiv F^c (\{\varepsilon_\ell\})$
with
$(\ell =2,3)$:
$$
F^{c} = \frac{1}{2}a_{1} \sum_{\vec{r},\vec{r}',\ell,\ell'}
\varepsilon_{\ell}(\vec{r},t){\it
U}^{c}_{\ell,\ell'}(\vec{r}-\vec{r}')\varepsilon_{\ell'}(\vec{r}',t)$$
$$ =
\frac{1}{2}a_{1}\sum_{\vec{k},\ell,\ell'}{\it
U}^{c}_{\ell,\ell'}(\hat{k})\varepsilon_{\ell}(\vec{k}, t)
{\varepsilon^*}_{\ell'}(\vec{k}, t).  \eqno(3.27b)$$

The Rayleigh dissipation function for the OP is

$$R=\frac{1}{2}\sum_{\vec{r},\ell}A'_{\ell}{\dot{\varepsilon}_{\ell}}^2,
\eqno(3.28a)$$
and the compatibility-induced contribution $\mathit{R^c}$ is the non-$OP$
dissipation
written in terms of the $OP$ strain rates

$$
{R}^{c} = \frac{a'_1}{2} \sum_{\vec{r},\vec{r}',\ell,\ell'}
\dot{\varepsilon}_{\ell} (\vec{r},t)
{\eta}^{c}_{\ell,\ell'}(\vec{r}-\vec{r}')\dot{\varepsilon}_{\ell'}(\vec{r}',t)
$$
$$=
\frac{a'_1}{2}\sum_{\vec{k},\ell,\ell'}{\eta}^{c}_{\ell,\ell'}(\hat{k})\omega^{2}
\dot{\varepsilon}_{\ell}(\vec{k},t)\dot{\varepsilon}_{\ell'}^*(\vec{k},t) . 
\eqno(3.28b)$$
Thus, explicitly
$$\frac{\delta ( R+R^c)}
{\delta \dot{\varepsilon_\ell}^* (\vec{k},\omega)}
= \sum_{\ell'} A'_{\ell,\ell'}
\dot{\varepsilon}_{\ell}(\vec{k},\omega), \eqno(3.28c)$$
where the effective $OP$ friction is
$$ A'_{\ell,\ell'} = A'_{\ell} \delta_{\ell,\ell'}
+ a'_1 {\eta^c}_{\ell,\ell'}. \eqno(3.28d) $$
Here the (frequency-independent) potential and  friction kernels
emerging from the dynamics are the
same,

$${\eta^c}_{\ell \ell'}(\hat{k}) = {\it {U}^c}_{\ell
\ell'}(\hat{k}), \eqno(3.29a)$$
where
$$ {U^c}_{\ell \ell'} = S_{1 \ell} S_{1 \ell'} = Q_{\ell,1}
Q_{\ell',1}\equiv Q_{\ell} Q_{\ell'} / {Q_1}^2,
\eqno(3.29b)$$
and (as from static-constrained minimization  \cite{Hatch}),
$$
{\it U}_{22}^{c}(\hat{k}) = \frac{(k_{x}^2-k_{y}^2)^2}{k^4} =
(Q_2/Q_1)^2;$$
$${\it U}_{33}^{c}(\hat{k}) = 4\frac{k_{x}^{2}k_{y}^{2} }{k^4} =
(Q_3/Q_1)^2;$$
$$ {\it U}_{23}^{c}(\hat{k})= \frac{2k_{x}k_{y}(k_{x}^2-k_{y}^2)}
{k^4} = Q_2 Q_3/{Q_1}^2= {\it U}_{32}^{c}(\hat{k}).
\eqno(3.29c)
$$

Both the friction and potential kernels depend on the wavector
direction $\hat{k}$, are independent of $|\vec{k}|$ for long wavelengths \cite{lattice},
and are `anisotropic long-range' functions encoding the
discrete symmetry of the lattice, that
fall off in coordinate space with a
dimensional power,
${\it \eta}^c (\vec{R}),{\it U}^c (\vec{R})\sim 1/R^D$.
At $\vec{k}=0$, the potentials are undefined and we set 
${\it U}_{\ell,ell'}^{c}=0$.
(A Coulomb potential, by contrast, diverges for long wavelengths as
$\sim 1/k^2$  and falls off as $\sim 1/R^{D-2}$.)

Although (3.26), derived from strain variation will give the same results
as (2.4), derived from displacement variation, they are
conceptually distinct.  In the
displacement picture, (2.8), or any equation obtained from it, is
solved with $(u_x,u_y)$ on a lattice, with both initial and boundary
conditions applied to the basic variables $\vec{u}$.  
Strains are defined as derivatives of the basic variables, and are 
derived quantities, e.g., $F \sim {\varepsilon_\ell}^4$ is a fourth power
of derivatives. By contrast, (3.26) is an $OP$-strain-only dynamics,
and is solved in the strain picture, with $\varepsilon_2,\varepsilon_3$
on a lattice; with initial and boundary conditions applied to the
basic variables $\{\varepsilon_\ell\}$; and with the $F \sim
{\varepsilon_\ell}^4$ term as zeroth order in derivatives. (Indeed,
the highest order derivatives are strain-gradient squared 
or Ginzburg terms.) The non-$OP$ strain
is a  derived quantity, obtained after solving
for the $OP$  and then using  (3.24), the displacement can also,
in principle, be derived  using (3.9).

In displacement-picture simulations, terms of the Landau free energy
and Rayleigh dissipation are all 
anisotropic, being powers of the various 
displacement-derivative combinations. In strain-picture simulations, the
anisotropy is in the
symmetry-specific  compatibility kernels, with $OP$ strains numerically
treated as  isotropic  
`scalars'. 
The strain picture has advantages,
as it
works directly with the physical strain variables, and uses compatibility
kernels evaluated once and for all, that encode the unit-cell symmetries
and give insight into energetically  favored textures.
\\

\noindent {\it SR underdamped dynamics:}

We now turn to the square-rectangle or $SR$ case, that shows a
different
structure, namely {\it time-retarded} $OP$ potentials and friction.
For the  $SR$ case , $n=2$, $N_{op}=1$, $N_c=1$; the
non-$OP$ strains are compression and shear $e_{1}$, $e_3$; the $OP$
strain is deviatoric $\varepsilon_2$; the symmetry constants are
$c_1=c_2={\sqrt{2}}$, $c_3=1$. The compatibility factors are
$Q_1({\vec{k}})= -{k^2}/\sqrt{2}$,
$Q_2=(k^2_x-k^2_y)/\sqrt{2}$, $Q_3=2k_xk_y$.
Then the compatibility constraint (3.6) becomes \cite{ourRC,Kartha}:
$$k^2e_1-{\sqrt{8}}k_xk_ye_3-(k^2_x-k^2_y)\varepsilon_2=0. \eqno(3.30)$$
The strain mass components are $\rho_{11}=\rho (k) (Q_1 /Q_3)^2 = 
\rho_{22}$,
$\rho_{12} =\rho (k) Q_1 Q_2 /{Q_3}^2 =
\rho_{21}$.
We have $n + N_{op} + N_c = 2+1+1= 4$ equations like (3.17), and
with arguments similiar to the $TR$ case we have, from the second part of
Appendix A, the equations (A9) for $\varepsilon_2, e_1, e_3$:

$$\rho_{22}\ddot{\varepsilon}_2+\rho_{21}\ddot{e}_1=-{\frac{\delta F}
{\delta\varepsilon_2^*}} -Q_2\Lambda -A_{2}'\dot{\varepsilon}_2,
\eqno(3.31a)$$
$$0 =-a_3e_3-a'_{3}\dot{e}_3 -Q_3\Lambda, \eqno(3.31b)$$
$$\{ \rho_{11}\ddot{e}_1+\rho_{12}\ddot{\varepsilon}_2\} =
-a_1e_1 - a'_{1}\dot{e}_1 -Q_1\Lambda, \eqno(3.31c)$$
$$Q_1e_1+Q_3 e_3=-Q_2\varepsilon_2. \eqno(3.31d)$$

Substituting for the compressional strain $e_1$ yields  coupled equations
between $e_3$ and $\varepsilon_{2}$. As in (3.23) it is convenient
to  define a
variable for the ubiquitous $Q$ ratios,
through $Q_{\ell ,\ell'} \equiv Q_{\ell}/Q_{\ell'}$. Then,

$$\rho_0 \ddot{\varepsilon_{2}} = -\frac{{c_2}^{2} \vec{k}^2}{4}
[ \
Q_{2,1}Q_{3,1}(a_{1} e_{3} + a'_{1} \dot{e}_{3})$$
$$+(a_1 Q_{2,1}^2 \varepsilon_2 +
F_{2}) + (a'_1 {Q_{2,1}}^2 + A'_2) \dot{\varepsilon}_{2} \ ],
\ \ (3.32a)$$
$$\rho_0 \ddot{e_{3}} = -\frac{ {c_3}^{2} \vec{k}^2} {4}
[ \ Q_{2,1}Q_{3,1}( a_{1}\varepsilon_{2}
+a'_1 \dot{\varepsilon_{2}})$$
$$+(a_{3} + a_1 {Q_{3,1}^2})e_3 +
(a'_{3} + a'_1 {Q_{3,1}^2})\dot{e_3}
\ ].
\eqno(3.32b)$$

Eliminating $e_3$, these yield the final result (A15) in terms of the $OP$
strain alone:
$$\rho_0\omega^2\varepsilon_2 =
\frac{{c_2}^2 k^2}{4} \left[ \frac{\delta F} {\delta\varepsilon_2^*}+
a_{1\omega}\frac {b_{\omega}(Q_{2}/Q_3)^2}{B_\omega}\varepsilon_2
- i\omega A_{2}'\varepsilon_2 \right] . \eqno(3.33)$$
As $b_\omega \equiv
{[a_{3\omega}-(4\rho_0 \omega^2/c_{3}^{2}k^{2})]}/{a_{1\omega}}$,
where $a_{i\omega} \equiv a_i - i \omega
a_{i}'$, the kernel is now frequency dependent, and complex,
as is the connection (A13) to  non-$OP$ strains:

$$e_1=S_{12}\varepsilon_2;  \ \ S_{12} \equiv
-\frac{(Q_{1}Q_2/{Q_3}^2)}{B_{\omega}}b_\omega ; \eqno(3.34a)$$

$$ e_3=S_{32}\varepsilon_2;  \ \ S_{32} \equiv  -\frac{(Q_2
/Q_3)}{B_\omega},
\eqno(3.34b)$$
where $B_\omega \equiv 1 + b_\omega (\frac{Q_1}{Q_3})^2$.
The compatibility condition is manifestly satisfied as an identity.

It is convenient to write the second term on the right-hand side of (3.33)
in terms of  real and imaginary parts
$$
\frac{a_{1\omega} (\frac{Q_2}{Q3})^2}{B_\omega}
\equiv { U}^{c}(\hat{k},\omega^2,(\frac{\omega}{k})^2)-
i\omega\eta^{c}(\hat{k},\omega^2,(\frac{\omega}{k})^2), \eqno(3.35)
$$
with ${ U^c}$, ${ \eta^c}$ given explicitly later. With this
separation,
$$
\rho_0\omega^{2}{\varepsilon_2}=\frac{{c_2}^2}{4}k^2 \left[ {{\delta
(F+F^{c})}\over
{\delta\varepsilon_2^{\star}(\vec{k},\omega)}}+{{\delta\left (
{R}+{R^c}\right)}
\over{\delta
\dot{\varepsilon}_2^{\star}(\vec{k},\omega)}} \right] \eqno(3.36)$$
where ${c_2}^2 = 2.$
The $OP$ dynamics for $\varepsilon_2 (\vec{r},t)$ is then
$$
\rho_0\ddot{\varepsilon}_2 (\vec{r},t)=
\frac{1}{2}{\vec{\Delta}^2 } \left[ \frac{\partial ( F+F^{c})}
{\partial{\varepsilon_2}(\vec{r},t)} + \frac{\partial
( \mathit{R}+\mathit{R^c} )}
{\partial
{\dot{\varepsilon}_2}(\vec{r},t)} \right], \eqno(3.37)$$

\noindent where the $OP$ triple-well free energy is the same
as in Appendix B:
$$F_0 =\sum_{\vec{r}} (\tau -1) {\varepsilon_2}^2 + {\varepsilon_2}^2
({\varepsilon_2}^2 -1)^2 +\frac{K_0}{2} (\vec{\Delta} \varepsilon_2)^2.
\eqno(3.38a)$$
Under four-fold rotations, $E_{xx} \rightarrow E_{yy}$, $E_{yy} 
\rightarrow E_{xx}$, $E_{xy} \rightarrow -E_{xy}$, the
non-$OP$ and $OP$ strains transform as $e_1 \rightarrow
e_1; \varepsilon_2 \rightarrow - \varepsilon_2; e_3 \rightarrow -e_3$,
leaving the free energy invariant and similarly for other 4mm point
group operations. The (anharmonic) $OP$ free energy $F_0$ is not
explicitly used in the derivation, and the dynamics depends only
on the number and type of the non-$OP$ strains.

We note that $\omega$ dependences carry an infinitesimal imaginary
part to maintain causality, and thus (3.37) has contributions only
from earlier times:
$$\frac{\partial F^c }{\partial \varepsilon_{2}} = \int_{-\infty}^{t}dt''
\sum_{\vec{r'}} U^{c}(\vec{r}-\vec{r}',t-t') \varepsilon_{2}(\vec{r}',t').
\eqno(3.38b)$$
The $OP$ dissipation in addition to
$${R}=\frac{1}{2}\sum_{\vec{r}}A'_{2}{\dot{\varepsilon}_2}^2,
\eqno(3.39a)$$

\noindent now also has a retarded compatibility contribution,

$$\frac{\partial R^c}{ \partial \dot{\varepsilon}_{2} }= 
\int_{-\infty}^{t} dt' \sum_{r'}
\eta^{c}(\vec{r}-\vec{r}',t-t')\dot{\varepsilon}_{2}(\vec{r}',t').
\eqno(3.39b)$$
Both are $ALR$ functions  ${\mathit U}^{c}$, $\eta^{c}\sim
1/R^D$ as before but  are  also {\it retarded in
time}. For negligible non-$OP$ friction  $a'_1 = a'_3 =0$, the
compatibility-induced friction vanishes, $\eta^c = 0$. The non-$OP$
compressional and
shear strains $e_{1,3} (\vec{r},t)$ are {\it
derived} quantities obtained after solving for $\varepsilon_2 (\vec{r},t)$
and using (3.34), and the constants $S_{i \ell}$ are frequency-dependent
and therefore retarded in time, like the compatibility potential.

The real and imaginary parts of the kernel of (3.35) that give rise to the
compatibility-induced $OP$ potential $U^{c}$ and the friction coefficient
$\eta^c$ are explicitly

$$U^{c}(\hat{k},\omega^2,v^2) =
\frac{a_3}{a_1}{Q_{2,3}}^2
\frac{ \left[1 + \frac{a_1}{a_3}|b_{\omega}|^2{{Q_{1,3}}}^2 \right] }
{ \left[ (1 +b_{r}{Q_{1,3}}^2 )^2 +
b^2_{i}{Q_{1,3}}^4 \right]
}, \eqno(3.40a)$$

\noindent where the `velocity' $v \equiv \omega/k$, and

$$b_{r} \equiv \frac{a_3}{a_1}\frac{ (g +
\omega^{2}\frac{a_{1}'}{a_1}\frac{a_{3}'}{a_3} ) }{( 1 +
(\frac{\omega a_{1}'}{a_1})^2)}, 
 \ \ b_{i} \equiv \frac{ \frac{a_3}{a_1}\omega{\left[g\frac{a'_{1}'}{a_1} -
\frac{a_{3}'}{a_3}\right]} }{ [
{1 +(\omega\frac{a_{1}'}{a_1})^2}]},$$

$$|b_{\omega}|^2 \equiv {b_r}^2 +{b_i}^2 = \left(\frac{a_3}{a_1}\right)^2
\frac{ \left(g^2
+
(\frac{\omega a_{3}'}{a_3})^2 \right) }{\left( 1 +
(\frac{\omega a_{1}'}{a_1})^2\right)}. \eqno(3.40b)$$

\noindent Similarly

$$\eta^c (\hat{k},\omega^2, v^2)=
\frac{
({Q_{1,3}Q_{2,3}})^4 |b_{\omega}|^2 +
\frac{a_{3}'}{a_{1}'}({Q_{2,3}})^4    }
{ \left[ (1 +b_{r}{Q_{1,3}}^2 )^2 +
b^2_{i}{(Q_{1,3})}^4 \right] },
\eqno(3.40c)$$

\noindent with

$$g \equiv 1- –\frac {4\rho_0}{c_{3}^2a_3} \left(\frac{\omega} {k}\right)^2.
\eqno(3.40d)$$

Note that both the static and dynamic  $U^c$ are zero for diagonal
orientations, when $Q_2 = 0$.
The zero-frequency limit $U^c (\hat{k}, 0, 0) \equiv {U^c}_0 (\hat{k})$ is
the static bulk compatibility potential  \cite{ourRC,Kartha}
used in earlier $TDGL$ simulations \cite{ourRC}, and is

$${U^c}_0 (\hat{k})= {Q_{2,1}}^2/ [ 1+(a_1/a_3){Q_{3,1}}^2 ].
\eqno(3.41a) $$
This favors  diagonal $k_x / k_y = \pm 1$ textures (when $Q_2 = 0$), and
the derived non-$OP$ strains $e_1,e_2$ are expelled in a kind of
`elastic Meissner effect' \cite{ourRC}, as can be seen from (3.34).
The bulk static compatibility potential is independent of $|\vec{k}|$ for 
long wavelengths and does not set a domain-wall separation length
scale. Near the center point ($\Gamma$) of the Brillouin Zone ($BZ$) the 
$U^c_0(\hat{k})$ depends only on the wavevector direction. 
It is
only in combination with the surface compatibility potential
\cite{surface}
$U^{surface}(k) \sim 1/k$ that equal-width `true' twins (that satisfy
a width-length scaling \cite{moore,bhk}) are
obtained \cite{ourRC}. 

An exact Fourier transform to coordinate space \cite{U(R)} confirms this
preference for diagonal orientations: with $\cos \theta =
\hat{r}. \hat{r}'$, the coordinate space compatibility potential
$${U^c}_0 (\vec{r}-\vec{r'})= \frac{G(\theta)}{|\vec{r}-\vec{r'}|^2}
\eqno(3.41b)$$
is found to have in the prefactor, a basic four-lobe structure 
from $\cos 4\theta$, with higher harmonics. 
With $a\equiv\frac{8a_1}{a_3}$, 

$$G(\theta) \equiv
\frac{[ 5 \cos4\theta(1 + \frac{\sin 2\theta}{5})
+ (17 + 4 \frac{a^2}{1+a^2}) \sin 2\theta
+ (13 - 4 \frac{a^2}{1+a^2}) ] }
{16{\sqrt{(1+a^2)}} (1 + \frac{a^2}{1+a^2}\cos 2\theta ) ^2}.\eqno(3.41c)$$

\noindent Figure 9 in Appendix C ($TDGL$ dynamics), with a single-site
initial condition, clearly shows multi-lobe strain textures from these
higher harmonics.

Returning to the dynamic case, we see from (3.40)  that the repulsive
kernel
has a {\it velocity-resonant structure} through $b_r \simeq g = 1 -
(v/v_0)^2$, strongest at a
propagation speed $ v =v_0 =
\sqrt{a_3/4\rho_0}$,
or a time-dependent propagation
length scale $L_p (t) =  t/t_p$ with $t_p = (4\rho_0 / a_3)^{1/2}$.
These non-$OP$ inertial effects, with anisotropic
directional
modulation or finite-velocity retardation
$\sim (\omega/k)^2$, compete with frictional delays  $\sim
\omega^2$, with the peaks becoming singular for vanishing
friction. This is somewhat like electrons interacting by
exchanging a photon, where the finiteness of the speed of light
produces retarded Coulomb potentials \cite{retard}. The resonant
structure   can be thought of as  the
$OP$ strain textures `exchanging a non-$OP$ phonon mode', causing inertial
time delays. $L_p (t)$ is
the time-dependent size of
an expanding  anisotropic `region  of influence' within which  changes in
texturing
at one point  enforce compatible changes in texturing at other points of
the lattice.
This phonon mediation is also described by the equivalent instantaneous
dynamics of (3.32)
with deviatoric and shear strains obeying two coupled underdamped
equations, that are convenient for numerical simulations \cite{Delay}.
\begin{figure}[h]
\epsfysize=6cm
\epsffile{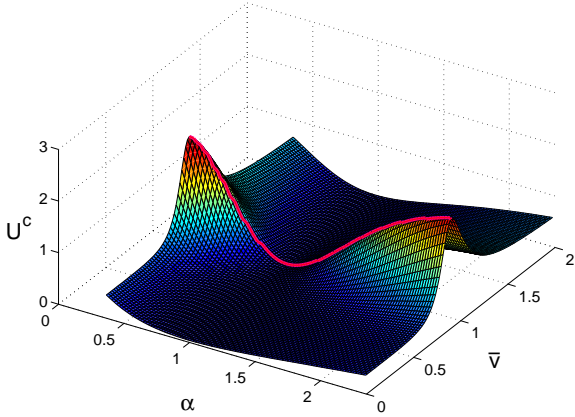}
\caption{Compatibility potential $U^c (\hat{k},\omega^2, (\omega/k)^2)$
for
the square-rectangle $SR$ case versus scaled velocity,
${\bar{v}} \equiv (4 \rho_0/a_3)^{\frac{1}{2}}(\omega/k)$, and
direction parameter $\alpha = k_x /k_y$
for frequency/dissipation parameter  ${\bar{\omega}}^2
=(\omega a'/a)^2 = 0.1$.
Thus $U^c$ is a repulsive and  dynamic orientational potential, favoring
diagonal textures, and moving interfaces.}
\label{fig2}
\end{figure}

The anisotropic potential kernel, 
$U^c(\hat{k},\omega^2,(\frac{\omega}{k})^2)$,
is plotted in Fig. 2. This shows plots of the kernel versus scaled
velocity,
namely ${\bar{v}} \equiv \sqrt{{\frac
{4\rho_0}{a_3} }}  \ v$  and various $BZ$ 
wave-vector directions \cite{lattice} $\alpha = k_x /k_y$ for scaled
frequencies $\bar{\omega}^2 \equiv \omega [(a'/a)]^2 = 0.1$, where we
take ${a'_1}/a_1
= {a'_3}/a_3 \equiv {a'}/a$. $U^c$ determines the positive energy costs of
the non-$OP$ strains, that the system wants to eliminate. The smallest
$U^c$ is at diagonal texturing  $|\alpha| = 1$, and zero velocity, $v =0$.
The most-non-optimal strains are the most strongly driven, with
large $U^c$. The striking features of Fig. 1 are clearly a preferred
diagonal
orientation $\alpha = \pm 1$, and a strongest-repulsion textural velocity
comparable to sound speeds, $\sim \sqrt{(a_3/4 \rho_0) [ 1 + a_1 /2 a_3
]}$ in the ferroelastic material at long times.
An extremal textural profile $\varepsilon_2 (\vec{k},\omega)$ in Fourier
space, if determined from an effective Lagrangian whose variation yields
(3.36), will be shaped by this peaked structure.
This suggests that textures will tilt diagonally and
thereafter remain stationary and rigid; and that the most unstable or
strongest-driven transient interfaces between phases will
move/grow at a constant velocity
close to the speed of sound, with bent domain-wall corners moving out,
or by segment tips growing,
along the  diagonals.

A fuller investigation of possible evolutions, with intermediate states
that
can be sensitive to elastic and frictional parameters,
requires further work.
Here we will only  illustrate the rich variety of texturings, and later
outline a
tentative scenario for nucleation and growth.
\begin{figure}[h]
\epsfxsize=7cm
\epsffile{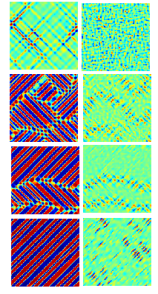}
\caption{Square-rectangle ($SR$) case: grain-boundary motion
under $BG$-type dynamics.
The columns show (top to bottom) temporal sequences for time $t$ of
10, 20, 42 and 70
`picoseconds´ (see Appendix B). The  initial conditions
are
$\varepsilon_2 (\vec{r},t=0), e_3 (\vec{r},t=0)$ random around
zero mean. Parameters (defined in Appendix B) are
$\rho_0=1,\tau=-0.25$, and the material is `hard', 
$a_1 =100, a_3 =210$; with ${A'}_2 =1, {a'}_3 = 0.1= {a'}_1$ . The time 
step is $\Delta t = 10^{-4}$.   Left column:
The $OP$ deviatoric strain
$\varepsilon_2 (\vec{r},t)$ under $BG$-equivalent dynamics
(3.32), showing formation of twin-like regions separated by grain
boundaries that are pushed out by domain-wall tip growth.
Right column: non-$OP$ shear strain $e_3 (\vec{r},t)$, that is
expelled from diagonal-domain regions, and concentrated in  the
pushed-out grain boundary regions ($e_1$
not shown).
Diagnostics \cite{simul} at $t=70$ were $E
= -1.08, \langle\varepsilon_2\rangle = 0.0012,$ max-min $\varepsilon_2
=(1.33,-1.35)$; $\langle e_3 \rangle = -7.9\times 10^{-6}$, max-min $e_3 =
(0.062,-0.058)$. 
}
\label{fig3}
\end{figure}

\begin{figure}[t]
\epsfxsize=7cm
\epsffile{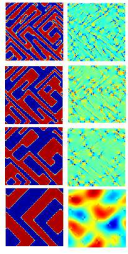}
\caption{Square-rectangle ($SR$) case: interface propagation under $BG$
type dynamics.
The columns show (top to bottom) temporal sequences for time $t$ in
picoseconds of 40, 80, 160 and 1000. The  initial conditions are
$\varepsilon_2 (\vec{r},t=0), e_3 (\vec{r},t=0)$ random around
zero mean. Parameters (defined in Appendix B) are
$\rho_0=1,\tau=-0.25$, and the material is `soft', $a_1 =10, a_3 =21$; 
with ${A'}_2 =1= {a'}_3, {a'}_1 =0$. 
The time step is $\Delta t = 0.002$. 
Left column: The $OP$ deviatoric strain
$\varepsilon_2 (\vec{r},t)$ under $BG$-dynamics 
(3.32), showing  domain walls propagating under
the repulsive long-range  compatibility potential, giving
the impression of a `zoom-lens' moving in. Right column: non-$OP$ shear
strain $e_3 (\vec{r},t)$, propagating outwards with interfaces,
concentrated at corners. ($e_1$ not shown.)
Diagnostics \cite{simul} at $t = 1000 $ were $E
= -1.4,\langle \varepsilon_2 \rangle = 0.0044$, max-min $\varepsilon_2
=(1.23,-1.23)$; $\langle e_3 \rangle = -0.0168$, max-min $e_3=(0.33,-0.49)$
.}
\label{fig4}
\end{figure}    
 
Figures 3 and 4 illustrate $SR$ case simulations \cite{simul} under the 
$BG$-equivalent dynamics of (3.32), with red/green/blue
representing positive/zero/negative strains, with relative color
intensities. Parameters are in the
captions and quenches are into the
martensitic regime $\tau = -0.25$.
Figure 3 shows time sequences of
$BG$ dynamics for the $OP$ deviatoric strain and the non-$OP$ shear
strain.  Note the grain-boundary-like regions that rapidly
anneal out by tip growth, carrying the shear strain that is already
expelled
from the diagonal-domain regions. For these parameters, the
propagation time $t_p < t_D $, the unit-cell relaxation
time defined later. Figure 4 shows the time
sequences, for smaller elastic constants (and hence smaller shear mode
velocities/larger propagation times), so $t_p > t_D$.  We now have
domain walls
propagating away from each other, giving the effect of a
`zoom-lens' moving in. Note the non-$OP$ shear fronts moving with the $OP$
walls (compare with Fig. 6 below). Such microstructure has been seen
in FePd using phase contrast microscopy \cite {sugiyama},
and twinning waves have been found in $1D$ models \cite{BG}.

\begin{figure}[h]
\epsfxsize=7cm
\epsffile{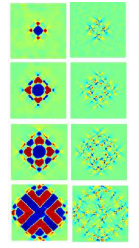}
\caption{$SR$ case: strain evolution under $BG$-type dynamics, with a fixed,
time-independent, Lorentzian-profile local stress. The  sequence  (top
to bottom), for time  in picoseconds of $t = 40, 60, 76, 106$
with the same parameters as Fig. 3,
but now $\tau = + 0.85$. Left column: Dynamic texturing of deviatoric 
strain
$\varepsilon_2 (\vec{r},t)$.
The system reduces the energy
from the imposed single-sign strain by `elastic photocopying', or
adaptive screening of the long-range elastic interaction,
generating higher multipoles  that
here
propagate. Right column: The non-$OP$ shear strain $e_3 (\vec{r},t)$
follows the $OP$ propagation.}
\label{fig5}
\end{figure}

\begin{figure}[h]
\epsfxsize=7cm
\epsffile{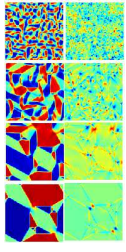}
\caption{Triangular to centered rectangle ($TR$) case: nested strain
texturing under $BG$ dynamics of (3.26).
The columns show temporal sequences, with time $t$ in picoseconds of
10, 40, 250, 620. The
initial conditions  for the $OP$ strains are
$\varepsilon_2 (\vec{r},t=0)$, $\varepsilon_3 (\vec{r},t=0)$ random around
zero mean. Parameters (defined in text) are $\rho_0=1,\tau=-50,
a_1 =1000, a_3 =2100; {A'}_2 =1, {a'}_3 = 1,  {a'}_1=0$ and the time step
is
$\Delta t = 10^{-3}$. Left column: The $OP$ shear strain
$\varepsilon_3 (\vec{r},t)$ showing
formation of nested star and triangle domains.
Right column: non-$OP$ compression
strain $e_1 (\vec{r},t)$, concentrated near domain corners,
and expelled elsewhere.
Diagnostics \cite{simul} at $t =620$ were $E
= -43.2$, $\langle \varepsilon_2 \rangle = -0.00316$,
$\langle \varepsilon_3 \rangle =
-1.05 \times 10^{-5}$, max-min $\varepsilon_2 =(4.45,-2.62)$, max-min
$\varepsilon_3 =(3.63,-3.65)$.}
\label{fig6}
\end{figure}

For parameters as in Fig. 4 and for an intermediate temperature, $\tau = 
0.85$,
Figure 5 shows dynamical stress responses to applied static 
deviatoric stress
$P_2 (r) = P_0 / [ 1 + (r/r_0 )^2 ]$, where  $r_0=1$ is the width
and $P_0 = 1$ is a time-independent strength.  The single-sign induced 
strain
has a large energy, and the system elastically screens it by nucleating
hierarchical opposite-sign elastic multipoles, with the propagation length
setting a scale, and wave fronts moving out. A sinusoidally time-varying
$P_0 (t)$ can produce even more striking propagating patterns.  
Similar `elastic photocopying'
was found previously under $TDGL$ dynamics, with the surface
compatibility potential \cite{ourRC,surface}
setting a domain-wall separation scale \cite{LANL,ourPhysica}

The last $BG$ simulation \cite{simul} of Fig. 6 shows, for the $TR$ case 
and (3.26), plots of the $OP$ shear $\varepsilon_2$, with star-triangle
patterns as found by other dynamics \cite{JacobsHO,Wen,ourAbstract}
and seen experimentally in crystals of lead orthovanadate, 
$Pb_{3}(VO_{4})_{2}$,
that undergoes a trigonal to monoclinic transition \cite{manolikas}.
See also Fig. 8 below.
Once again, the $e_1$ plot shows that equilibration involves expelling
non-$OP$ strains (except at $OP$ corners, with  $e_1$ global 
cancellation).

Finally, we note that linearizing the $BG$  dynamics (3.36)
about equilibrium in the zero-damping limit
yields the familiar wave equation.
`Textural phonon' spectra can emerge.  For
a finite  $L_0 \times L_0$ system we
can also include surface-compatibility restoring forces, with a
kernel
\cite{ourRC}, $U^{surface} \sim (a_3/a_1L_0) /|\vec{k}|$.
The long wavelength $OP$ strain  oscillations then have  velocities
$\omega (\vec{k})/k = v(\hat{k})$ that are obtained by solving
$$ v^2 \sim \frac{1}{2\rho_0} [ <F''> + a_1 U^c  + a_1
U^{surface}],  \eqno(3.42)$$
\noindent where $<F''>$ is a free energy curvature averaged with a
probability
distribution peaked at the equilibrium structure.
For infinite systems the long wavelength spectrum is  linear. For finite
$L_0$, very long wavelengths probe the system size \cite{bhk}, and
$\omega \sim (|k|/L_0)^{\frac{1}{2}}$. This is the ``dyadon´´ spectrum
\cite{bhk} of waves in twin-bands of martensite. Anomalies
have indeed been observed \cite{Anom} in some ferroelastic phonon spectra.
However, we do not pursue this conjectured explanation here.
\subsection{ External stress and noise terms }

We now consider stress $\{ p_s(\vec{r})\}$
and  delta-correlated noise $\{ \tilde{g}_
{s}(t) \}$, that modify the free energy as

$$F \rightarrow F -
\sum_{\ell,\vec{r}}(p_{\ell}\varepsilon_\ell+\tilde{g}_{\ell}\varepsilon_\ell)
,$$
$$f \rightarrow f -\sum_{i,\vec{r}}(p_{i}e_i+\tilde{g}_{i}e_i),
\eqno(3.43a)$$
where  noise correlations are 
$$<\tilde{g}_{\ell}(\vec{r},t)\tilde{g}_{\ell'}(\vec{r}',t')> =
2 A'_\ell \bar{T}
\delta_{\ell ,\ell'}\delta_{\vec{r},\vec{r'}}\delta
(t -t');$$
$$<\tilde{g}_{i}(\vec{r},t)\tilde{g}_{i'}(\vec{r}',t')> =
2 a'_i \bar{T}
\delta_{i ,i'}\delta_{\vec{r},\vec{r'}}\delta
(t -t'), \eqno(3.43b)$$
i.e. the bare noise is Markovian. Here
$\bar{T} \equiv k_B T /E_0$, and $E_0$ is an $OP$ elastic energy
of Appendix B.
For small stress and noise, a simple use of the substitution (1.4),
justified by a detailed analysis, yields
effective $OP$ stresses and noises,
$$ {p_{\ell}}^{tot} = p_{\ell} + {p_{\ell}}^{c}; \  {\tilde{g}_{\ell}}^{tot}
= \tilde{g}_{\ell} + {\tilde{g}_{\ell}}^{c}, \eqno(3.44a)$$
where
$$ {p_{\ell}}^{c}= \sum_i p_i S_{i \ell};\ \
{\tilde{g}_{\ell}}^{c} = \sum_i \tilde{g}_i S_{i \ell}, \eqno(3.44.b)$$
with constants $S_{i \ell}$ as in (3.24) and (3.34), respectively for
the $TR$ and $SR$ cases.
The total correlations are
$$<\tilde{g}^{tot}_{\ell} (\hat{k},t) {\tilde{g}*^{tot}_{\ell'}}
(\hat{k'},t')> = 2 \bar{T} 
{A'}_{\ell,\ell'}(\hat{k})\delta_{\vec{k},\vec{k'}}
\delta (t-t'), \eqno(3.44c)$$
where as in (3.28d)
$${A'}_{\ell,\ell'}(\hat{k}) \equiv  \delta_{\ell,\ell'}A'_{\ell}
  + \sum_{i}  a'_i S_{i \ell} S^{*}_{i \ell'}. \eqno(3.44d)$$

The elimination of non-OP strains
thus induces cross-couplings, so non-$OP$ stresses induce $OP$ variations;
noise correlations become spatially
nonlocal; and different $OP$'s acquire cross-correlated noises.
The $BG$ deterministic dynamics then becomes the $BG$ Langevin dynamics of
(1.3).
For the $TR$ case the noise is delta-correlated in time. For the
$SR$ case with $OP$ only, there is frequency-dependence, but this can be
circumvented by considering  $\{\varepsilon_2, e_3\}$ of (3.32) as our
system.
Thus in both cases we have two variables with Markovian noises.

\subsection{ Fokker-Planck Description }

Langevin dynamics with delta-correlated Markovian noise can be
\cite{FP}
equivalently written in  a Fokker-Planck ($FP$)
description. The set of ${4 {L^2}_0}$ random variables labelled by
$\alpha = \{\ell,\vec{k}\}$ is taken to be

$$ \{ x_\alpha \} = \{ \varepsilon_2 (\vec{k},t) ,
\varepsilon_3 (\vec{k},t) \};
\{v_\alpha\}=\{ \dot{\varepsilon}_2 (\vec{k},t), \dot{\varepsilon}_3
(\vec{k},t)\}.  \eqno(3.45)$$
\noindent for the $TR$ case, with $\varepsilon_3 \rightarrow e_3$ for the
$SR$ case.
The Langevin equations are
$$\dot{x}_{\alpha} (t) = v_{\alpha} (t),$$
$$\dot{v}_{\alpha} (t) =
{D^{(1)}}_{\alpha} + \hat{\Gamma}_\alpha (t) ,  \eqno(3.46a)$$
where the frictional force plus internal stress, or  `drift' term is
$${D^{(1)}}_{\alpha} = -\frac{1}{M(k)} \frac{\partial (F + F^c)}{\partial
{x_{\alpha}}^*}
- \sum_{\alpha'} {D^{(2)}}_{\alpha,\alpha'} v_{\alpha'} , 
\eqno(3.46b)$$
where $M(k)=4\rho_0/k^2$ is a strain mass. The Langevin noise 
correlation is
$$ <\hat{\Gamma}_\alpha (t) \hat{\Gamma}_{\alpha'} (t')> =
2 {D^{(2)}}_{\alpha,\alpha'}  \delta (t-t'),
\eqno(3.46c)$$
with  `diffusion coefficient'
$${D^{(2)}}_{\alpha,\alpha'} = \delta_{\vec{k},\vec{k'}}
\bar{T} {A'}_{\ell,\ell'}(\hat{k})/M(k)^2 .  \eqno(3.46d)$$

The $FP$ equation in Kramers form \cite{FP} for the time-dependent 
probability $P(\{ x_\alpha \},\{v_\alpha \}, t)$ is

$$ \partial P /{\partial t} = \hat{L} P \equiv
[\hat{ L}^{(1)} + \hat{ L}^{(2)}] P,  \eqno(3.47a) $$

\noindent where the Fokker-Planck operator for ferroelastics is a sum of 
drift and diffusion terms, respectively, given by

$$ \hat{ L}^{(1)} = -v_{\alpha}\frac{\partial}{\partial {x_{\alpha}}^*}
-\frac{\partial}{\partial {x_{\alpha}}^*} D^{(1)}_{\alpha},$$
$$\hat{ L}^{(2)} = \sum_{\alpha,\alpha'}\frac{{\partial}^2}{\partial
{x_{\alpha}}^*\partial
{x_{\alpha'}}^*} {D^{(2)}}_{\alpha,\alpha'}.
\eqno(3.47b)$$

A formal solution   for the probability in terms of the initial distribution
is \cite{FP}

$$ P(\{ x_\alpha \},\{ v_\alpha \} ,t) = e^{ t \hat{ L}} P (\{ x_\alpha
\}),\{ v_\alpha \}, 0).
\eqno(3.48) $$

\noindent The $FP$ operator carries the nonlinearity, symmetries,
anisotropies,
and long-range spatial correlations. Its eigenvalues and eigenfunctions
can be used to describe dynamic correlations. Since `potential
conditions' hold \cite{FP}, the asymptotic probability is
a Boltzmann distribution, that (as can be checked by substitution) is
$ P_0 = e^{-[(F + F^c) + \frac{1}{2} \sum_\alpha M {v_\alpha}^2]/\bar{T}}$.
The free energy minima thus correspond to probability peaks
in $OP$ function space, highest for zero strain rate. For uniform $OP$
this implies
a triple-well free energy, but for nonuniform textures there
will be a
more complex free energy landscape. The multiple extrema are the
$TDGL$ asymptotic states.

The $FP$ formalism is convenient for discussing  textural
dynamics, metastability and glassy behavior; e.g. to
determine strain correlations that correspond to experimentally
probed response functions \cite{Anantha,planes}, or to calculate
temperature-dependent transformation
rates through first-passage times \cite{Madan}.
Both strains and strain-rates appear naturally, as in phenomenological
models of elasticity and plasticity \cite{Bala}, that could thereby be
given a
microscopic basis.

\section{BG Dynamics  as an Inhomogeneous  Array of
Damped and Coupled Nonlinear Oscillators}

Here, we consider a mechanical analog of nonidentical damped
oscillators, that suggests a physical scenario for nucleation and growth
after a temperature quench.
We first review well-known damped oscillator results, to fix notation and
terminology, and regimes of validity.

A particle of mass $M$ and friction parameter $A$, driven by a
force $ -F'\equiv - \partial F/\partial x = -A x$ of spring constant $A
\equiv
F''$,
obeys the
underdamped equation

$$M\ddot{x} + A' \dot{x} = -F' , \eqno(4.1a)$$
that  can be written in an equivalent convenient form

$$\ddot{x} = -(1/M) [ \partial F/\partial x + A' \dot{x} ]
, \eqno(4.1b) $$

\noindent and the general solution is oscillations of exponentially
decreasing
amplitude. With natural frequency ${\omega_0}^2 = F'' /M$ and relaxation
rate
${\tau_0}^{-1} \equiv A' /M$, the complex frequency is \cite{osc}

$$\omega ' = \sqrt{\omega_0^2 - (\tau_0)^{-2}} -i \frac{1}{2}
\tau_0^{-1} . \eqno(4.1c)$$

\noindent Thus there is exponential decay (without even one complete
oscillation) in the overdamped parameter  limit of

$$ \tau_0 {\omega_0} < 1 , \eqno(4.2)$$

\noindent when the inertial term is small at all times, and

$$\dot{x} (t) \approx -\lambda [\partial F/\partial x(t)] ,
\eqno(4.3a) $$

\noindent with  $\lambda = 1/A'$, describes the damped behavior. 
Outside this regime it  approximately describes the
exponentially decaying envelope of the oscillations,
at low frequencies/long times:

$$\omega << A'/M; ~~~\  t >> 2\pi M/ A'.
\eqno(4.3b)$$

\noindent Note that from (4.1c), and (4.3b), larger
frequency/lower mass oscillations are damped out earlier.

We now turn to $BG$-type evolution equations, that cannot be
obtained by adding simple local inertial terms to
the standard  overdamped
dynamics like Model A or B dynamics for a non-conserved
or conserved order parameter, respectively \cite{HH}.
However, defining an inverse mass ${1/M(k)} =
1/\rho (k) =k^2 /4\rho_0 \sim k^2$ the apparently unusual $BG$ structure
of (3.26) for the $TR$ case, say, can be written as
$$\ddot{\varepsilon_{\ell}} (\vec{k},t) = - (1/M(k))
[ \partial (F + F^{c})/\partial {\varepsilon_{\ell}}^* (\vec{k},t)
+ A'_{\ell} \dot{\varepsilon_{\ell}} (\vec{k},t)], \eqno(4.4a) $$
\noindent where for simplicity, $a_{1}'=0$.
On comparing with (4.1b), the dynamics has an intuitively
appealing interpretation.
It is the dynamics of a set of (nonlinear, coupled) oscillators,
of (two-component) spring extension \cite{complexosc}
$\varepsilon_{\ell}(\vec{k}, t )$, labeled by $k$, with different
masses $M(k) \sim
1/k^2$ that are strongly dependent on the oscillator label
$k$: heaviest near the origin, and lightest at the Brillouin-zone
($BZ$) edge. Similarly, the damping rate ${\tau_0 (k)}^{-1} \equiv
A'/ M(k) \sim k^2$ is smaller for the larger masses.
The spring coupling $U^c (\hat{k})$ acts equally over all the $BZ$ 
oscillators, in a
given direction $\hat{k}$. Note that there is an intrinsically
large range of damping times $\tau_{0} \sim k^{-2}$, over orders of magnitude.
(This is reminicent of decay times at criticality at a second order critical point).
From (4.4a) the $k \rightarrow 0$ infinite-mass oscillator
is a special case, with its initial
velocity $\dot{\varepsilon} (k \rightarrow 0,0)$ unchanged
\cite{neutrality}.

\begin{figure}[h]
\epsfxsize=7cm
\epsffile{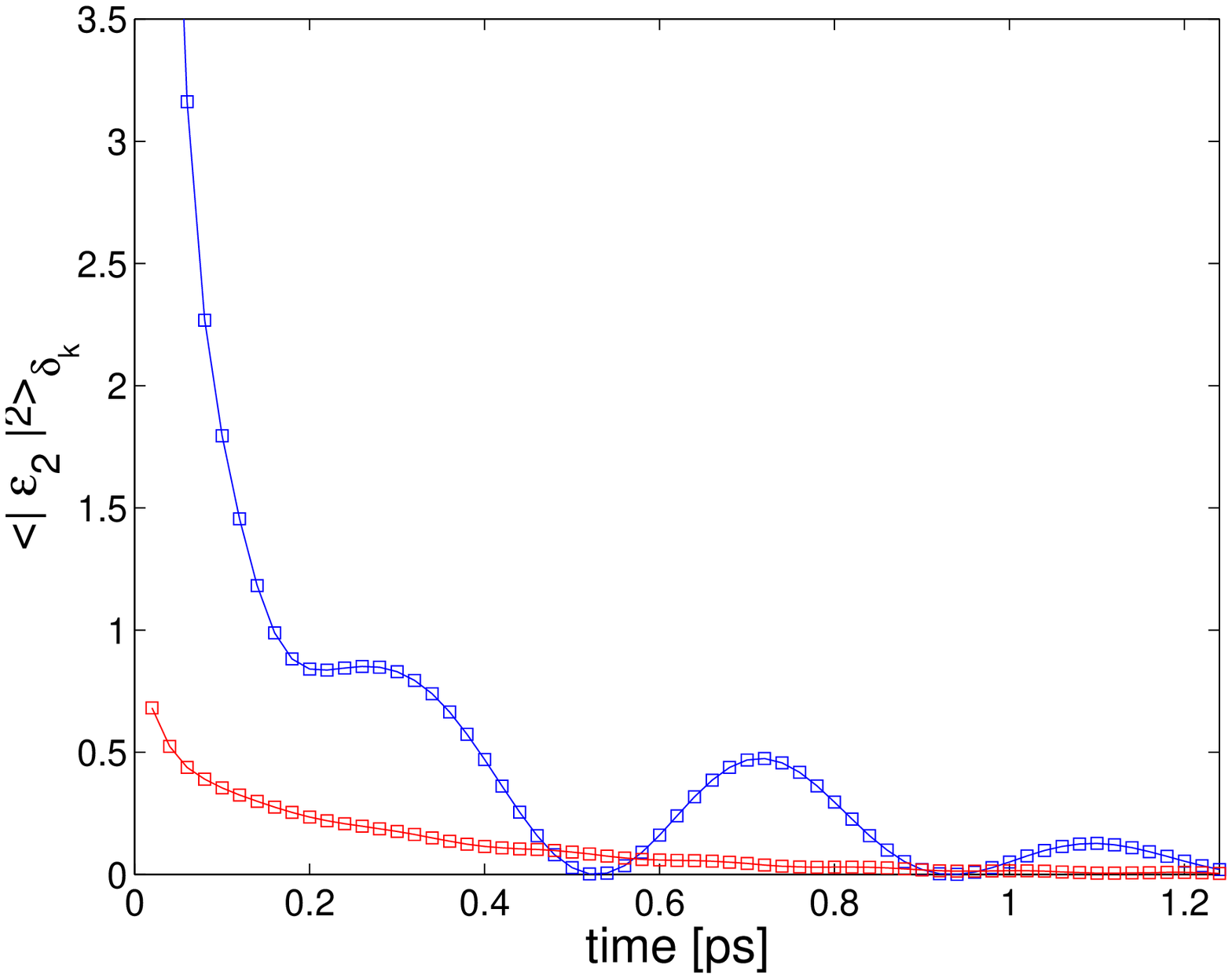}
\epsfxsize=7cm
\epsffile{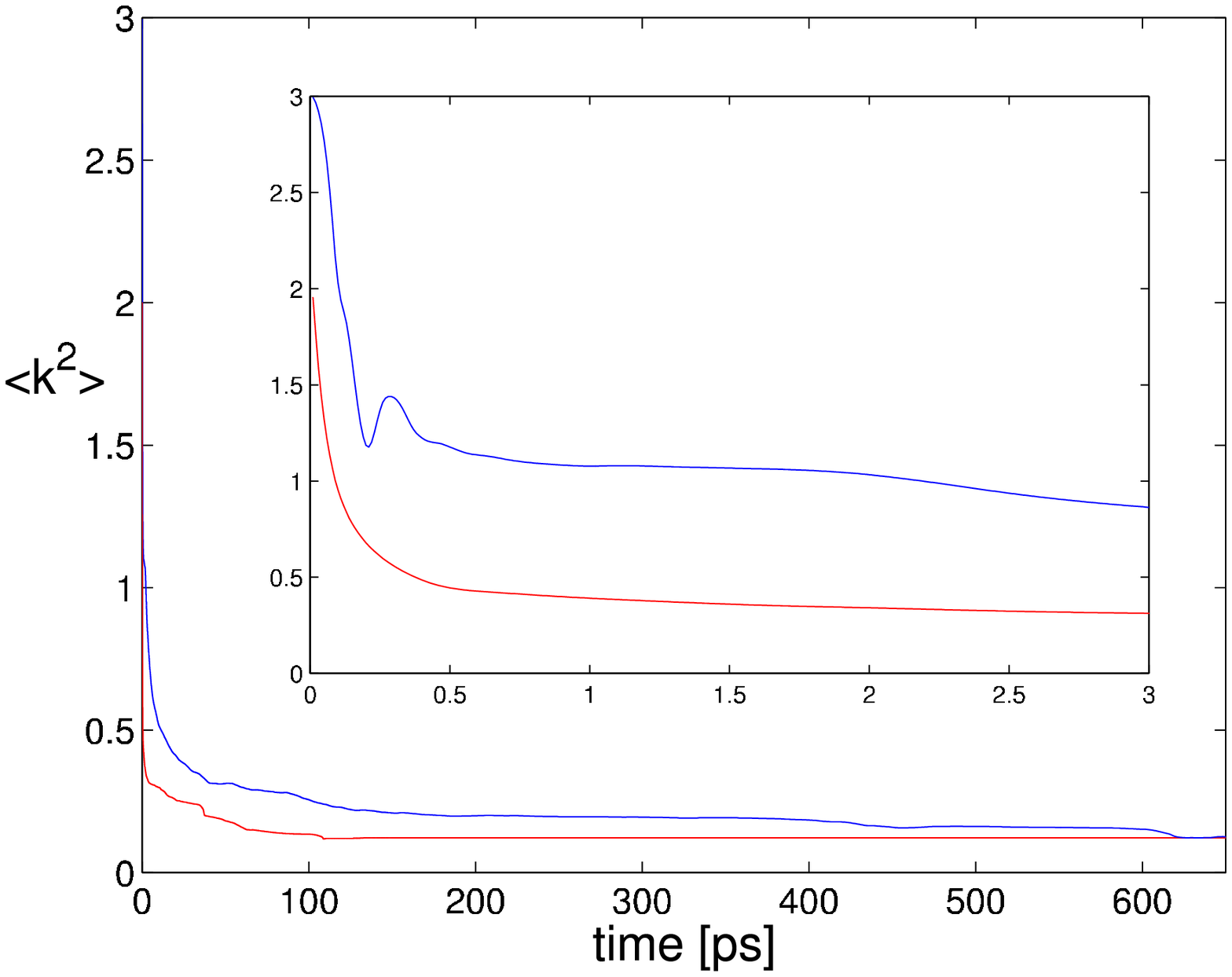}
\caption{{\it Top:} Angularly averaged square oscillator extension or
strain structure factor $\bar{S}(k,t)$ (defined in the text) versus 
time in picoseconds; for $SR$-case nonlinear $BG$-dynamics oscillators, 
labeled by k. The oscillators are at high temperatures $\tau=4$ 
(where there is only a single-well free energy), with parameters as 
in Fig. 4, and an initial condition as in text.  The set of inhomogeneous 
strain oscillators of different mass $\sim k^{-2}$ and damping $\sim k^2$, 
are averaged over a shell of thickness $2 \delta k$ in the Brillouin zone. 
The heavier, less damped oscillators with smaller $k =0.525 $ 
(blue) oscillate, while the lighter, strongly damped oscillators with $k =
2.65 $ (red) are overdamped. The shells are $2\delta=0.05$ and $0.1$, 
respectively.   
{\it Bottom:} Comparison of $<k^2>=\sum_{k}
k^2 |\varepsilon_{3}|^2/\sum_{k}|\varepsilon_{3}|^2$ for
$BG$ and $TDGL$ dynamics for  TR oscillators. The $BG$ line (blue) 
asymptotically merges with the $TDGL$ line (red).  Inset shows early time
damped oscillations in the $BG$ case and overdamped monotonic decay for 
the $TDGL$ case.  
}
\label{fig7}
\end{figure}


Linearizing around equilibrium extensions, with effective  curvature
$F'' = {F_0}'' + K_0 k^2 \equiv A(k)$, and with ${\omega_0}^2 \equiv
A(k)/M(k)$, the
complex frequency is
$$\omega ' =  \sqrt{{\omega_0 (k)}^2 - (\tau_0 (k))^{-2}} -i \frac{1}{2}
{\tau_0 (k)}^{-1}. \eqno(4.4b)$$

The top panel of Fig. 7 shows that the angularly averaged strain structure 
factor
$\bar{S}(k,t) = < |\varepsilon_2 (\vec{k},t) |^2 >_{\delta k}$ or squared oscillator
extension averaged over a shell $2\delta k$, is underdamped for small-$k$ oscillators,
but is overdamped for
larger-$k$ oscillators. The bottom panel shows asymptotic agreement between
$BG$ and $TDGL$ dynamics. (See also Figs. 6, 8). 
The initial strains are nonzero only in a $5 \times
5$ square region, and the oscillating strains show up as oscillating
colors (not shown). The bottom panel shows that for the $SR$ case the moment
$<k^2>$ takes on asymptotically the same value in the $BG$ and $TDGL$ dynamics.

From (4.4b), the long wavelength modes are of
necessity underdamped, with $\omega_0 \tau_0 \sim 1/k >> 1$, a point made by
Reid and Gooding \cite{Reid}. We will for simplicity
consider the regime where {\it none} of the oscillators  are in the
overdamped regime (4.2),
i.e. $ \tau_0 (k)\omega_0 (k) >> 1$ for all $k$, even those at the
$BZ$ corner $k =\sqrt{2} \pi$.
The oscillating, damped texture has to settle down to {\it some}
time-independent state,
that from Sec. III.D  will be a peak of the probability distribution,
where the
free energy derivative is zero.
Then an `envelope dynamics'  as in (4.3a)
for the $OP$ strain is
$$\dot{\varepsilon}_{\ell} (\vec{k}, t) = - \lambda \frac{\partial
(F+F^{c})}{\partial {\varepsilon_{\ell}}^* (\vec{k}, t)},
\eqno(4.5a)$$
where $\lambda = 1/A_{\ell}'$.
However, the different-mass/friction oscillators can reach this late-time
behavior only at different times $t_k$.
Thus for times
and lengths introduced through $\omega \sim 2 \pi /t$, $k \sim \pi /L$,
there is a wavevector-dependent time scale $t > t_k$ or a
time-dependent length scale $L< L_D(t)$ for textures to achieve the
late-time regime, namely
$$\omega << A'/M(k) ; ~~~\ t > t_k \equiv   2 \pi /[(A' /4\rho_0)k^2]
,$$
$$ L < L_D (t) = (t/ t_D)^{1/2}; ~~~\  t_D \equiv (4\rho_0/\pi A')^{1/2},
\eqno(4.5b)$$

\noindent where $t_D$ is the time for relaxation across a unit cell.

Taking the idea  of a late-time equilibration  length  $L_D (t)$ as more
generally
applicable to the nonlinear case, a tentative picture
emerges.
While damped harmonic oscillators have trivial and identical final states, an
inhomogeneous set of nonlinear
oscillators with long-range coupling can have $\vec{k}$-dependent
inhomogeneous final states
$\{\varepsilon (\vec{k}, t \rightarrow \infty )\}= \{\bar{\varepsilon}
(\vec{k})\}$.
The oscillations begin around the average value (say, zero) of the
initial states $\{\varepsilon (\vec{k}, t =0 )\}$,for times $t >t_D$ 
after a temperature quench.
Because of the $k$-dependent inhomogeneous damping  of
(4.4a), the
lightest masses near the $BZ$ corner, with labels $\pi> k > \pi /L_D (t)
\equiv k_D (t) $ will be the first to
feel the final state attractor and begin oscillations around
the final rather than the initial state. As time proceeds, the boundary 
shifts,
and
the circle of $k < k_D (t)$ `initial-state' oscillators shrinks, while
the number of $\sqrt{2} \pi > k > k_D (t)$ `final-state' oscillators
increases.
Finally, all except the very smallest-$k$ oscillators are on an
equilibration path, and the larger-$k$ ones (corresponding to structures of 
time-dependent
size $L_D (t)$) 
 have already reached
it.
There is thus {\it sequential-scale equilibration}, from the edge of the
$BZ$ inwards.
For the $SR$ case \cite{Delay}, the  
additional $L_p (t)$ propagation length enriches the scenario.
The tentative scenario is consistent with what we have seen in our
simulations for given parameters, although the nonlinear
mode-coupling could induce more complex relaxation pathways,
in more general parameter regimes.
 
 Thus {\it the unusual  $BG$
dynamics implies unusual elastic properties.} Since $\rho (\vec{k}) \sim
1/k^2$, long
wavelength strains are kinematically
blocked from decaying too early, and the scale-dependent damping and
equilibration process starts at small-scale textures oriented by
compatibility potentials, and then spreads
to larger length scales, with associated non-$OP$ strain
expulsion. Consequently, materials classes governed by $BG$
dynamics can
have rich spatial patterning, metastability and glassiness, hierarchical
multiscale microstructure,
and a complex nonequilibrium elastic response.

\section{LATE TIME/SMALL SCALE LIMITS OF DYNAMICS}
We  now show (for all frictions $a'_i$, $A'_{\ell}$ nonzero) the 
underdamped $TR$ and $SR$  equations are approximated by  
$TDGL$-type  equations. The validity of $TDGL$ dynamics in
${\it any}$ regime was recently questioned \cite{JacobsHO} and an alternative 
overdamped dynamics, obtained by dropping displacement-acceleration 
terms in (2.8) was proposed \cite{JacobsHO,JacobsSR}. Appendix C shows 
they are equivalent. \\

\noindent{\it TR case dynamics:}

For the $TR$ case, the late-time equations, from dropping  $OP$
strain-acceleration terms in (3.26), are with (3.28c),

$$ \sum_{\ell'} A_{\ell, \ell'}(\hat{k}) \dot{\varepsilon}_\ell (\vec{k}, t)
= - \frac{\partial
(F+F^{c})}{\partial\bar{\varepsilon}_{\ell}^{\star}(\vec{k},t)}
. \eqno(5.1)$$

\noindent This truncation is valid only for textures in a low 
frequency/large wave vector 
regime as in (4.5b). For $a'_1=0$ (4.5a) follows, while for the 
general case, an inversion yields
$$
\dot{\varepsilon_\ell}
(\vec{k},t)=-\sum_{\ell'}\lambda_{\ell\ell'}(\hat{k})
\frac{\partial (F+F^{c})}{\partial\varepsilon_{\ell'}^{\star}(\vec{k},t)},
\eqno(5.2a)$$
where the Onsager coefficient matrix is, with $A'_2 =A'_3$
\[
\b{\b{$\lambda$}}
=
\frac {-1}{\cal W}
\left[
\begin{array}{cc}
A_{3}'+a_{1}'Q_{3,1}^2      &  -a_{1}'Q_{2,1}Q_{3,1}  \\
-a_{1}'Q_{2,1}Q_{3,1}      &  A_{2}'+a_{1}'Q_{2,1}^2
\end{array}         \right],  \ \ \ \ \ \ \ \ \ \ \ \ \ \ \ \ (5.2b)
\]

\noindent with the determinant (actually isotropic for $A_{2}'=A_{3}'$)
$${\cal W} =A_{2}'A_{3}'\left[ 1+(\frac{a_{1}'}{A_{3}'})Q_{3,1}^2+
(\frac{a_{1}'}{A_{2}'})Q_{2,1}^2 \right]. \eqno(5.2c)$$

Since diagonal elements, and the determinant, of the $\lambda$ matrix are
nonzero, (5.2) implies that asymptotic textures
${\bar{\varepsilon}}_{\ell} (\vec{k})$ are
determined by the extrema of the total free energy,
that locate stable and metastable textural minima, as in the $FP$
discussion of Sec. III.D:

$$\partial
(F+F^{c})/\partial\bar{\varepsilon}_{\ell}^{\star}(\vec{k})
=0.  \eqno(5.3)$$

\noindent In coordinate space,
$$\dot{\varepsilon_\ell}
(\vec{r,t})=-\sum_{\ell'}\sum_{\vec{r}'}\lambda_{\ell\ell
'}
(\vec{r}-\vec{r}')
\frac{\partial (F+F^{c})}{\partial\varepsilon_{\ell'}(\vec{r}',t)}.
\eqno(5.4)$$
This is a  time-dependent Ginzburg-Landau
equation, with
Onsager coefficients that are anisotropic and
spatially {\it nonlocal}.
For negligible non-$OP$ friction parameter ($a_{1}'/A'_2 \rightarrow0$), 
the Onsager coefficient
matrix becomes both spatially independent and diagonal in $OP$ labels, 
yielding an `ordinary' $TDGL$ equation with a constant,
isotropic, and uniform friction:
$$ \dot{\varepsilon_\ell} (\vec{r},t)=
-\frac{1}{A'_\ell} \frac{\partial
(F+F^{c})}{\partial\varepsilon_{\ell}(\vec{r},t)},
\eqno(5.5)$$
but still anisotropic and nonlocal in the elastic forces.\\

\noindent{\it SR case dynamics:}

For the $SR$ case, from dropping $OP$ inertial terms in (3.36) in
$\vec{k},\omega$ space,
$$F_{2} -i\omega [A_{2}' +
\eta^{c}(\hat{k},\omega^2,v^2) ]\varepsilon_{2} +
{\mathit U}^{c}(\hat{k},\omega^2,v^2)
\varepsilon_{2} =0.  \eqno(5.6a)$$
This can be written as a generalized  $TDGL$-type equation with
{\it retarded} Onsager kernels $\lambda (\vec{r} - \vec{r'}, t-t')$
in coordinate space
$$\dot{\varepsilon}_2
(\vec{r},t)=-\int_{-\infty}^{t} dt'\sum_{\vec{r}'}\lambda
(\vec{r}-\vec{r}', t-t')
\frac{\partial (F+F^{c})}{\partial\varepsilon_2 (\vec{r}',t')}.
\eqno(5.6b)$$
Since the general $TDGL$ structure is a `current' proportional to a
`force', the Onsager coefficient $\lambda(\vec{k}, \omega) \equiv 1/[ A'_2
+ \eta^c (\hat{k},\omega^2,v^2)]$ is like a dynamic
`conductivity'. We consider however, the
low frequency $\omega\rightarrow 0$, or asymptotic
long-time limit, keeping the linear term in frequency, but neglecting the
quadratic and higher order $\omega^2, (\omega/k)^2$ frequency
dependence in the kernels. The regime of validity is considered below.
This yields  an instantaneous  nonlocal $TDGL$ equation as in the $TR$
case, (5.4), with a single $\ell=2$ order parameter. 
In Fourier space,
$F^{c}={1\over 2}\sum_{\vec{k},\omega}{{\mathit U}^{c}}_0
(\hat{k})
|\varepsilon_2
(\vec{k} ,t)|^2 $,
where
the retarded compatibility
potential reduces to the static expression \cite{ourRC,Kartha}
of (3.41) 
and the Onsager coefficient $\lambda (\hat{k})$ is given by
$\lambda (\hat{k}) \equiv  1/[A'_2 + \eta^c (\hat{k},0,0)]$.

For  non-$OP$ friction $a'_{1,3}/a_{1,3} \rightarrow 0$, $\eta^c$
vanishes, and we again recover
an ordinary (local, instantaneous) $TDGL$ equation as in (5.5) and $\ell=2$ 
with compatibility forces remaining nonlocal. Thus the model used in
previous work \cite{ourRC} is a specific limit of the exact
dynamics.

In a regime similar to (4.5b) (with $d'^2 = Max[ (a'_1 a'_3/a_1
a_3), (a'_1/a_1)^2] $ ), the  non-$OP$ inertial delay
$(\rho_0/a_3)(\omega/k)^2$ and frictional retardation
$ \sim (\omega d')^2$  can be respectively
neglected, yielding (5.5), for lengths  $ L <
Min[L_D (t), L_p (t)]$,
and times $t > {t}_f \equiv (2\pi  d')^{1/2}$. These  simple heuristic
estimates may not, of course, be strictly
quantitative, but capture the diffusive aspect of the late-time
relaxation.
\begin{figure}[h]
\epsfxsize=7cm
\epsffile{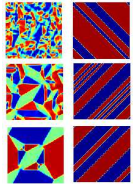}
\caption{Ordinary $TDGL$ dynamics and $OP$ textures:
Left column: evolution for $TR$ case with time $t=1,20,150 $ and parameters
$a_1 = 1000, \tau = -50 ,A'_2 =
A'_3 = 1$. Right column: $SR$ case, $a_1 = 100, a_3 = 210, \tau = -0.25
, A'_2 = 1$ showing states evolved from different initial conditions, but
the same parameters, illustrating nearly-degenerate multiple
free energy minima discussed in the text.}
\label{fig8}
\end{figure}
Figure 8, left column, shows that the $TDGL$ dynamics \cite{GL} for the $TR$
case
yields textures similar to the (longer-run) $BG$ dynamics of Fig. 6.

We note that the free energy can have several metastable minima with different
microstructure, but closeby free energy densities. Thus the nested stars of 
Fig.6, under perturbation, yield rhombhohedral structures of slightly lower energy
density \cite{icomat02} (that are also obtained directly with a different
initial random-number seed). The multiple-minima picture also emerges
in $TDGL$ simulations
for the $SR$ case of Fig.8 where different random-number seeds
produce 
three $SR$ case quasi-twin textures, with diagonal-domains
of different number and separation \cite{delta}. Such quasi-twins were
previously obtained in a displacement picture by Monte Carlo simulation
\cite{Kartha}. The total free energies
in each case are extensive $\sim {L_0}^2$, while their
energy differences behave as the length $\sim K_0 L_0$  of the diagonal
domain
walls, (with  compatibility cost from $U^c$ vanishing): the free
energy density difference is then $\sim 1/L_0$.
(A surface compatibility potential sets a domain-wall separation
length scale \cite{ourRC} and would raise the degeneracy, favoring equal-width
`true' twins.) The barrier between
states with differing numbers of walls is the cost $\sim A_1 L_0$ of a
fractional-length  kink in the domain wall and the barrier crossing time
$\sim e^\frac{{A_{1}L_{0}}}{\bar{T}}$.

Thus, in general the $ALR$ potential can produce a
multiple-minima free energy landscape, with nearly degenerate 
, differently-textured states
separated by
large barriers. Initial conditions or intermediate-state dynamical scales
can lock the system into one of the metastable 
states.
The possibility of multiple minima 
is consistent with recent analyses of models of competing
short and long range interactions \cite{schmalian}.

In general, the inclusion of noise in $BG$-Langevin simulations will
enable the system to more easily find low-energy minima.
We now turn to other  symmetries, and give the
dynamics for the four crystal systems  of $2D$
ferroelastic transitions.

\section{OP STRAIN DYNAMICS FOR ALL 2D-SYMMETRY TRANSITIONS}

Since the derivations above involved the generic strain mass
tensor and the harmonic non-$OP$ strains  and  did not involve the
details of the anharmonic $OP$ free energy, the $BG$ dynamics structure will be
the same for all types of $2D$ (and indeed, $3D$) transitions. However,
the nature of the non-$OP$ strains determines the symmetry of
the compatibility kernels.

In $2D$ there are $N=3$ strains (compressional, deviatoric and shear) and
one compatibility equation $(N_c=1)$. The distinct ferroelastic 
transformations (Fig. 1),
order parameters and all possible symmetry-allowed combinations of
strains in the Landau free energy, were found by Hatch et al. \cite{Hatch}
using the computer program ISOTROPY. These  transformations
fall into  two classes determined by the nature and number of order
parameters.  (A) For either a two-component $OP$ or two one-component $OP$'s
($N_{op}=2, $ $n=1$), we have  the $TR$ case  and the square to oblique
($SO$) case, respectively.  (B) For a one component $OP$, ($N_{op}=1, n=2$),
we have the $SR$ case with deviatoric
$OP$; the square to centered rectangle ($SC$) case with shear $OP$; and
the rectangle to oblique ($RO$) case with shear $OP$.
In group theoretical
symmetry notation, the transformations in Fig.1 are: (a) P4mm to P2mm, (b)
P2mm to P2, (c) P4mm to C2mm, (d) P6mm to C2mm, (e) P6mm to P2, and (f)
P4mm to P2. The $OP$ compatibility kernels are frequency independent (dependent)
for case A (B). 

\subsection{One non-$OP$ strain, two  $OP$ strains  \
($N_{op}=2$, $n=1$):}

\begin{itemize}
\item $TR$ case, driven by combined deviatoric and shear strains
$\varepsilon_2,\varepsilon_3$.
This is studied in the text above.

\item $SO$ case, driven by independent deviatoric $\varepsilon_2$ and
shear strains, $\varepsilon_3$.
For the $SO$, there are two distinct $OP$'s  that drive the transition,
and the
$OP$ free energy is \cite{Hatch}

$$F_0 =\frac{1}{2} \sum_{\vec{r}} A_2 \varepsilon_2^2+
B_2\varepsilon_2^4+A_3\varepsilon_3^2+
B_3 \varepsilon_3^4, \eqno(6.1) $$
\end{itemize}
where the constants are merely illustrative.
However, the harmonic non-$OP$ energy and thus the form of the
dynamics and the kernels, are identical in both cases.\\\

\subsection{Two non-$OP$ strains, one $OP$ strain \
($N_{op}=1$, $n=2$):}

\begin{itemize}
\item {\it $SR$ driven by deviatoric strain, $\varepsilon_2$}.
This is considered in detail in the text above.

\item {\it $SC$ driven by shear strain, $\varepsilon_3$}
The $OP$ free energy now is as in (3.38a) but is $F_0(\{\varepsilon_3\})$.
The non-$OP$ harmonic energy, $ f=\frac{1}{2}\sum_{r}a_{1}e_{1}^2 +
a_{2}e_{2}^2$ , and the dissipation is
$ R^{tot}=[{\frac{1}{2}} \sum_r a'_1\dot{e_1}^2
+ a_{2}'\dot{e_2}^2+ A_{3}'\dot{\varepsilon_3}^2]$.
Thus the derivation carries over, with the interchange $2\leftrightarrow
3$, and, in particular, the compatibility factors and symmetry
constants interchanged,
$Q_2 \leftrightarrow Q_3, c_2 \leftrightarrow c_3$ in the
dynamical kernel of (3.40).
The $\omega \rightarrow 0$ limit, analagous to (3.41), is the static result. 

$$a_1{U^c}_0 (\hat{k})= \frac{a_1{Q_{3,1}}^2}{ [ 1+(a_1/a_2){Q_{2,1}}^2 ]}.
\eqno(6.2) $$
This is zero for $k_x =0$ or for $k_y =0$, so domain walls are
either vertical or horizontal, as can be confirmed in simulations.
The dynamics for $e_2, \varepsilon_3$ are:
$$\rho_0 \ddot{e_{2}} = -\frac{{c_2}^{2} \vec{k}^2}{4}
[ \
Q_{2,1}Q_{3,1}(a_{1} \varepsilon_{3} + a'_{1} \dot{\varepsilon}_{3})$$
$$+(a_1 {Q_{2,1}}^2 +a_2)e_2  + (a'_1 {Q_{2,1}}^2 + a'_2) \dot{e}_{2} \
], \eqno(6.3a)$$
$$\rho_0 \ddot{\varepsilon}_{3} = -\frac{ {c_3}^{2} \vec{k}^2} {4}
[ \ Q_{2,1}Q_{3,1}( a_{1} e_{2}
+a'_1 \dot{e}_{2})$$
$$+( F_3 + a_1 {Q_{3,1}^2}e_3 ) +
(A'_{3} + a'_1 {Q_{3,1}^2})\dot{\varepsilon}_3 + X
\ ], \eqno(6.3b)$$
where the extra term $X \equiv 0$ in (6.3b).

\item {\it $RO$ driven by shear strain, $\varepsilon_3$}.
The non-$OP$ free energy is harmonic in the sum and difference of the
compression and deviatoric strains \cite{Hatch}:

$f=\frac{1}{2}\sum_{r}\{a_{+}(e_{1}+e_{2})^2 + a_{-}(e_{1}-e_{2})^2\}$

$ \ \ \ =\frac{1}{2}\sum_{r}\{a_{1}{e_{1}}^2 + a_{2}{e_{2}}^2 + 2a_X e_1
e_2 \},$
\noindent where $a_{1} \equiv a_{+} + a_{-} \equiv a_2$ and $a_X \equiv
a_{+} - a_{-}$.
Similiarly
$ R^{tot}=\frac{1}{2}\sum_r [A'_3{\dot{\varepsilon}_3}^2
+ a_{1}'{\dot{e_1}}^2+ a_{2}'
{\dot{e_2}}^2 + 2 a'_X {\dot{e}}_1{\dot{e}}_2]$.
The derivation carries through.  The static limit compatibility kernel is:
$$a_1{U^c}_0 (\hat{k})= {Q_{3,1}}^2 \left(\frac{{\bar{a}}_1}{[
1+({\bar{a}}_1/{\bar{a}}_2){Q_{2,1}}^2 ]} + a_X Q_{1,2} \right).
\eqno(6.4)$$
Here, ${\bar{a}}_1 \equiv a_1 - a_X Q_{1,2};
{\bar{a}}_2 \equiv a_2 - a_X Q_{2,1}$. The dynamics is as in (6.3),
but with the substitution $a_{1,2} \rightarrow {\bar{a}}_{1,2}$,
and $a'_{1,2} \rightarrow {\bar{a}'}_{1,2}$, where
${\bar{a}'}_1 \equiv a'_1 - a'_X Q_{1,2}; 
{\bar{a}'}_2 \equiv a'_2 - a'_X Q_{2,1}$. The extra term in (6.3b)
is $X \equiv (a_X \varepsilon_3 + a'_X \dot{\varepsilon}_3 ) {Q_{3,1}}^2
Q_{1,2}$.
This again shows that vertical/horizontal domains are favored,
but now the relative magnitudes of the original non-$OP$
elastic constants $a_{\pm}$ act like symmetry-breaking fields, with $x$
axis orientation preferred for $a_{X} >0$,
and $y$ axis preferred for $a_{X} < 0$.  The predicted behavior
can be verified in simulations.

\end{itemize}

\section{SUMMARY AND DISCUSSION}

There are two themes in this paper: firstly, a derivation of
ferroelastic evolution equations for all $2D$ symmetries, and secondly,
a demonstration of a strain-based (rather than a
displacement-based) description of elastic phase transitions.

We have, for the first time, derived the $D >1$ underdamped dynamics
for ferroelastics in terms of the order-parameter ($OP$)
strains $\{\varepsilon_\ell\}$ alone,
showing that the evolution equations are of a generalized Bales-Gooding
form. The strain-based derivation yields  a wavevector
dependent strain mass
$\sim 1/\vec{k}^2$, thus large-scale strains have greater inertia.
The structure is that of an $OP$ strain acceleration,
$\ddot{\varepsilon_\ell}$,
proportional to a Laplacian acting on the sum of an $OP$-only stress and an
$OP$-only frictional force. The stress and friction are  strain
and strain-velocity derivatives, respectively, of the
effective free energy $F(\{\varepsilon_\ell \}) +
F^{c}(\{\varepsilon_\ell \})$ and
effective
Rayleigh dissipation $R(\{ \dot{\varepsilon_\ell} \}) +
R^{c}(\{ \dot{\varepsilon_\ell} \})$.
These contain, in addition to direct local $OP$ contributions
($F$ and $R$), additional anisotropic and long-range contributions ($F^c$ 
and
$R^c$) that
emerge from eliminating the non-$OP$ strains using St. Venant's
compatibility conditions. The kernels are explicitly evaluated for
all 2D symmetries. There are also compatibility-induced noise
contributions, and this $BG$-Langevin
dynamics of (1.3) or (3.46) is a central result. A Fokker-Planck 
equation (3.47) is obtained. The $BG$ dynamics can be regarded as 
nonlinear, nonlocally coupled
oscillators labeled by $\vec{k}$,
with unequal masses $\sim 1/\vec{k}^2$, and dampings $\sim \vec{k}^2$. The 
textures are the set of final
rest positions $\{ \varepsilon_\ell (\vec{k},t \rightarrow \infty) \}$, with large-$k$
oscillators (small-scale strain textures) equilibrating first.
The late-time envelope dynamics that guides the damped oscillations to
this equilibration  is of the $TDGL$ form.
This analog suggested an  appealing
picture of sequential-scale evolution
for post-quench nucleation and hierarchical growth, accounting for
nonuniform textures.

We adopt the strain picture in simulations, with strains as the basic
variables on sites of a reference lattice, driven by symmetry-allowed
terms of the Ginzburg-Landau free energy, and by anisotropic, 
symmetry-specific,
long-range compatibility forces.
The free energies are in a standardized
form, with dimensionless parameters related to experiment.
Simulations show that the $BG$ dynamics has rich texturing properties,
including
repulsive velocity-resonant compatibility potentials that can drive
interfaces at (nearly) sound speeds.

We now place our results in perspective with some of the other models that
have earlier provided valuable insights. Baus and
Lovett \cite{Baus} invoked the $19$th century work of St. Venant
\cite{venant,compat} on the
compatibility condition for the strain tensor, in the context of
surface tension
in liquids. They considered strain as the basic variable in the argument,
and noted it might be useful in elastic solids.
We similiarly work in the
strain picture, differing in this respect from previous
\cite{Anantha,Madan,Kartha} simulations that work with
$\vec{u}$ gradients, i.e. in the displacement picture.

In an interesting and important paper, Kartha et al. \cite{Kartha}
performed Monte Carlo simulations to find static textures. They
used the $SR$ free energy  that is sixth power in the deviatoric strain
order parameter and harmonic in the compressional and shear non-OP
strains, and the simulations were in terms of the displacement, so
effectively  $V=V(\{\vec{u}\})$. Since striking textures, like
unequal-separation diagonal domain walls (and tweed) were obtained, 
they attempted to understand these $\vec{u}$ simulation textures by using
compatibility \cite{compat,Baus} to eliminate non-OP strains in the
free energy, plotting (static) compatibility potentials. However, these
effective strain-strain correlations were not directly used
in the simulations. Such explicit implementation of the compatibility
forces was done in a $TDGL$ strain dynamics, where quasi-twins, the elastic
Meissner effect (expulsion of non-OP strains) and tweed were investigated
\cite{ourRC}. Other $TDGL$ displacement simulations investigated 
tweed alone \cite{Onuki}. (The tweed
terms
considered \cite{ourRC,Onuki,Kartha} were all
different.)

Our work has the same Lagrange-Rayleigh starting point
\cite{BG,LL} as Refs. \onlinecite{JacobsHO,JacobsSR}, that
focus early on in their argument on an
overdamped-displacement dynamics. We follow a
different path and
derive an OP-only underdamped strain
dynamics, finding  it is of a generalized $BG$ form; and that the
late-time limit is a $TDGL$-type equation.

Our approach differs from an underdamped dynamics \cite{Anantha} for
the $SR$ case, that did not explicitly consider non-$OP$ strains; and
phenomenologically added a static anisotropic long-range potential
between squares of strains to explain acoustic signals \cite{planes}.
We derive the $BG$ structure from a Lagrangian with both non-$OP$
strains and compatibility constraints, and find a
retarded anisotropic long-range force in terms of the $OP$ strains
themselves (and not their squares). Our dynamics in the $TDGL$
limit also differs from a $TDGL$ dynamics \cite{Khatch,Wen},
where strains have been eliminated in favor of morphological profile
variables, $\eta_{\alpha}(r,t)$, with ${\alpha}$ labeling
the structural variants. The static  potentials between squares of
the morphological variables were obtained from 
elastic fields due to inclusions \cite{eshelby}.

Our approach is in the spirit of the Landau description of phase
transitions \cite{Landau}: working with the order parameters as the
basic and physically relevant variables, and focusing on the order
parameter symmetries (as encoded in the compatibility factors),
as the source of  ferroelastic static and dynamic
texturing.

Further work could include a detailed understanding of $2D$
nucleation, growth and interfacial profiles; extensions to $3D$
symmetries such as cubic to tetragonal \cite{our3DPRL} ($N_{op}=2$, 
$n=4$, $N_c=6)$; generalizations to include defects, in a broader
`strain elastodynamics' framework; 
making contact with phenomenologies of plasticity; 
simulations
and calculations of experimentally measurable strain correlations and
nonlinear susceptibilities; and exploring a hierarchical scenario for
shape memory \cite{ourPhysica}.

The symmetry-specific, compatibility-focused underdamped
ferroelastic dynamics for the strain order parameter encode, in their
very structure, the possibility of an evolutionary textural hierarchy in
both space and time, and a tendency for interfaces to be driven at
sound speeds, explaining some of the
fascinating but puzzling features of martensitic dynamics.
The dynamical equations could be applied to a wide variety of
textural evolutions that include improper ferroelastics,
leading to a deeper understanding of many  materials
of technological interest such as ferroelectrics, magnetoelastics,
colossal magneto-resistance manganites, superconducting cuprates, and
shape memory materials.

\section{Acknowledgment}
It is a pleasure to  acknowledge stimulating discussions with Professors
G.R. Barsch, S. Franz, Y.B. Gaididei, D.M. Hatch, K. Kawasaki and J.A.
Krumhansl.  TL is grateful to The University of Western Ontario (UWO) and
NSERC of Canada for support. SRS thanks The Centre for Chemical Physics at 
UWO
for a Senior Visiting Fellowship and the Theoretical Division,
LANL for hospitality. This work was supported by the U.S. Department of 
Energy.

\appendix

\section{SQUARE TO RECTANGLE TRANSITION DYNAMICS}

We need to demonstrate explicitly that the same dynamics results whether
displacements or strains are treated as the basic independent variables.
We derive, in two self-contained subsections
the (same) underdamped
dynamics for the $SR$ case, (a) by  varying the displacement, and (b) by
varying the strains subject to compatibility.

\subsection{ Variation of displacements}

The $(n=2)$ non-$OP$ strains are here the compressional $(e_1)$ and shear
strains $(e_3)$, and the $N_{op}=1$ deviatoric strain
$(\varepsilon_2)$ is the $OP$:
$$\frac{e_1}{c_1}={1\over  2}\left(\Delta_x u_x+\Delta_y u_y\right),
~~\frac{\varepsilon_{2}}{c_2}={1\over  2}\left(\Delta_x u_x-\Delta_y
u_y\right),$$
$$\frac{e_{3}}{c_3}={1\over 2}\left(\Delta_x u_y+\Delta_y u_x\right),
\eqno(A1)$$
with the free energy $V=f+F$, where
$$
f=\frac{1}{2}\sum_{r}(a_{1}e_{1}^2 + a_{3}e_{3}^2). \eqno(A2a)$$
\noindent The Rayleigh dissipation function is
$$
{ R}^{tot}=\frac{1}{2}\sum_r [a'_1{\dot{e}_1}^2
+ a_{3}'{\dot{e}_3}^2+A_{2}'{\dot{\varepsilon}_2}^2]. \eqno(A2b)$$
From (A2) and varying with respect to $\vec{u}$ as in (2.1) and (2.2) ,
$$
\rho_0\ddot{u}_x=  {1\over 2}\{\Delta_x [f_{1}+{ R_1} ] +
\Delta_y[f_{3}+{ R_3}] + \Delta_x[G_{2}+{ R_2}]\},
\eqno(A3a)$$
$$
\rho_0\ddot{u_y}=  {1\over  2}\{\Delta_y[f_{1}+{ R}_1] +
\Delta_x[f_{3}+{ R}_3] - \Delta_y[G_{2}+{ R}_2]\}.
\eqno(A3b)$$

This is the result of Ref. \onlinecite{JacobsSR}, where

$$f_{1,3} \equiv c_{1,3}\frac{\partial f}{\partial e_{1,3}};
~~G_{2} \equiv c_2 \frac{\partial F}{\partial
\varepsilon_{2}};$$
$$ { R}_{1,3} \equiv c_{1,3} \frac{\partial {
R}^{tot}}{\partial \dot{e}_{1,3}}; ~~{ R}_{2} \equiv c_2 \frac{\partial {
R^{tot}}}{\partial
\dot{\varepsilon}_{2}}. \eqno(A3c)$$
The underdamped strain equations with
$\vec{\Delta}^{2}\equiv\Delta_{x}^2+\Delta_{y}^2, 
\hat{D^2}\equiv\Delta_{x}^2-\Delta_{y}^2$, are:
$$
\rho_0\ddot{e}_{1} = \frac{c_1}{4}[{\Delta}^2(f_1 + R_1)
+ 2\Delta_{x}\Delta_{y}(f_3 + R_3)$$
$$ + \hat{D}^2  (G_{2}+ R_2)],
\eqno(A4a)$$
$$
\rho_0 \ddot{\varepsilon}_{2} = \frac{c_2}{4}[+\hat{D}^2(f_1 + R_1)
+\vec{\Delta}^2 ( G_{2}+ R_2 )],
\eqno(A4b)$$
$$
\rho_0 \ddot{e}_{3} = \frac{c_3}{4}[2 \Delta_{x}\Delta_{y}
(f_1 + R_1) + \vec{\Delta}^2(f_3 + R_3)].
\eqno(A4c)$$
We also have the compatibility relation of (3.6)
$$ \hat{Q}_1 e_1  + \hat{Q}_2 \varepsilon_2 + \hat{Q}_3 e_3=0, \eqno(A5)$$
where $\hat{Q}_1 ={\vec{\Delta}}^2 /c_1, \hat{Q}_2 =-\hat{D}^2/c_2 ,
\hat{Q}_3= -2 \Delta_x \Delta_y /c_3$.
By taking derivatives of (A4) we see that (A5) is identically satisfied.
Thus
we can take (A4a),(A4c) and (A5) as the three equations to determine the
three
strains, since then two of the equations are linear.
Fourier transforming these three equations, we obtain
for $e_{1,3}(\vec{k},\omega)$, $\varepsilon_2(\vec{k} ,\omega)$
$$
\rho_0\omega^2\varepsilon_{2} =
\frac{1}{4}[c_1 c_2 (k_{x}^2-k_{y}^2)a_{1\omega}
e_{1}
+  {c_2}^2 k^2( F_{2} - i\omega A_{2}'\varepsilon_{2})],
\eqno(A6a)$$
$$
\rho_0 \omega^2e_{3} = \frac{1}{4}[c_1 c_3 2 k_{x}k_{y} a_{1\omega}e_{1}
+ {c_3}^2 k^2 a_{3\omega} e_{3}],
\eqno(A6b)$$
$$ Q_1 e_{1} = -Q_3 e_{3} - Q_2 \varepsilon_{2},
\eqno(A6c)$$
\noindent where $F_{2}=\partial
F/\partial {\varepsilon_{2}^{\star}(\vec{k}
,\omega )}$; $a_{1\omega}\equiv a_{1} - i\omega a_{1}';
  a_{3\omega}\equiv a_{3} - i\omega a_{3}'$  and $Q_1 = k^2 /c_2 ,
Q_2 = (k_x ^2 - k_y ^2)/c_2, Q_{3}=2 k_x k_y /c_3$.

Defining
$b_{\omega}\equiv [a_{3\omega} - \rho_0
(\frac{2\omega}{k})^2]/a_{1\omega}$ and $B_{\omega} \equiv 1 +
b_{\omega} (Q_1 /Q_3)^2$, we obtain

$$e_{1}(\vec{k},\omega) = (Q_1 /Q_3) b_{\omega} e_3(\vec{k},\omega),$$
$$e_{3}(\vec{k},\omega)= -(Q_2/Q_3 B_\omega) \varepsilon_2
(\vec{k},\omega).
\eqno(A7)$$
Using (A6), and (A7)
$$
\rho_0\omega^{2}\varepsilon_{2}=\frac{{c_2}^2}{4}k^2[F_{2} - i \omega
A_{2}'
\varepsilon_{2} +  a_{1\omega} (Q_2 /Q_3)^2 (b_\omega
/B_\omega)\varepsilon_{2}].
\eqno(A8)$$
The dynamics is written out in $BG$ form and discussed at the end of Sec. III.

\subsection{ Variation in Strain}

By varying (3.1) with respect
to $e_{1,3}({\vec{r}},t)$, $\varepsilon_2({\vec{r}},t)$, we obtain
$N+N_c=3+1=4$ equations:

$$\rho_{22}\ddot{\varepsilon}_2+\rho_{21}\ddot{e}_1=-{\frac{\delta F}
{\delta\varepsilon_2^*}}-Q_2\Lambda -A_{2}'\dot{\varepsilon}_2,
\eqno(A9a)$$
$$0 =-a_3e_3-a'_{3}\dot{e}_3 -Q_3\Lambda, \eqno(A9b)$$
$$\{ \rho_{11}\ddot{e}_1+\rho_{12}\ddot{\varepsilon}_2\} =
-a_1e_1 - a'_{1}\dot{e }_1 -Q_1\Lambda -a'_3 \dot{e_3}, \eqno(A9c)$$
$$Q_1e_1+Q_3\varepsilon_3=-Q_2\varepsilon_2 . \eqno(A9d)$$
\noindent In $({\vec{k}},\omega )$ space for $e_{1,3}({\vec{k}},\omega )$,
$\varepsilon_ 2({\vec{k}},\omega )$ and with $a_{1\omega}\equiv a_1-i\omega
a_{1}$,
$a_{3 \omega}\equiv a_3-i\omega a_{3}$, there is one nonlinear equation,
$$\omega^2\rho_{22}\varepsilon_2 +\omega^2\rho_{21}e_1=+{\frac{\delta F}
{\delta\varepsilon_2^*}}+Q_2\Lambda -i\omega A_{2}'\varepsilon_2
\eqno(A10)$$
and $n+N_c=3$ linear equations
$$0 =-a_{3\omega}e_3-Q_3\Lambda, \eqno(A11a)  $$
$$-\omega^2\left(\frac{\rho_0}{{c_3}^2 c_1 Q_3}\right)e_3=
-a_{1\omega}e_1+ Q_1\Lambda =0, \eqno(A11b)$$
$$Q_1e_1+Q_3e_3=-Q_2\varepsilon_2 , \eqno(A11c)$$

\noindent where we have used a relation as in (3.21).  From (A11a) we have
the Lagrange multiplier
$$\Lambda ({\vec{k}},\omega )=-\frac{a_{3\omega} e_3}{Q_3},  \eqno(A12)$$
and hence the non-$OP$ strains in terms of each other,
$e_1=+b_{\omega}(\frac{Q_1}{Q_3})e_3$. Finally in terms of the $OP$,
$$e_3=-(Q_2/Q_3){\varepsilon_2}/B_\omega, \eqno(A13a)$$
where
$$b_{\omega} \equiv \frac{[a_{3\omega}-\rho_0 (2\omega
/k)^2]}{a_{1\omega}}; \ \
B_\omega \equiv  1 + b_\omega (Q_1 / Q_3)^2. \eqno(A13b)$$
Then
$$\frac{4 \rho_0\omega^2}{{c_2 ^2} k^2}\varepsilon_2=+ \left[
F_2  - i\omega A_{2}'\varepsilon_2
+W \varepsilon_2
\right], \eqno(A14a)$$
where
$$W \equiv  \frac{4\rho_0\omega^2}{{c_2}^2 k^2} -\omega^2 [\rho_{22}
-\frac{Q_1
Q_2}{{Q^2}_3} \frac{b_{\omega} \rho_{21}}{B_{\omega}} + a_{3 \omega}
\frac{(Q_2 /Q_3)^2}{B_{\omega}}]. \eqno(A14b)$$
We use the definitions (3.12) of $\rho_{22}=\rho_{11}$ $=(4 \rho_0 /{c_3 ^2}
k^2 ) (c_1 Q_1/c_3 Q_3)^2$; $\rho_{21}=(4 \rho_0 / k^2 {c_3}^2 )(Q_1 Q_2
/{Q_3}^2)  \rho_{12}$ and
$({Q_1}^2 - {Q_2}^2)/{Q_3}^2 = (c_3/c_2)^2$$ = 1/2$ to obtain:
$$
\rho_{0}\omega^{2}\varepsilon_{2}=\frac{{c_2}^2}{4}k^2[F_{2} - i \omega
A_{2}'
\varepsilon_{2} + a_{1\omega}(Q_2 /Q_3)^2 (b_{\omega} /B_{\omega})
\varepsilon_{2}]. \eqno(A15)$$
Comparing with (A8), we see that the same dynamics results, whether
displacement or strain is regarded
as the independent variable, with compatibility enforced in 
the
latter case. The $BG$ evolution equation is discussed at the end of
Sec. III.

A useful intermediate equation, obtained by substituting $e_1$ in  
terms of $e_3$ into (A10) and using (A12) for the Lagrange multiplier,
expresses the equations as a coupled set for the $OP$ and non-$OP$
shear. This dynamics  is given in (3.32), and is entirely
equivalent to the
$OP$-only retarded equation of (A15).

\section {SCALING OF FREE ENERGY}

Here we generalize the free energy scaling of Barsch and Krumhansl
\cite{Fscaling} to scale the ${\it Lagrangian}$ such that the Landau free energy
is in a standard polynomial form; all the parameters are
dimensionless; strains are of order unity and times are scaled in a
characteristic
time unit.

The symmetric strain tensor is related to displacement derivatives through
$$\phi_{\mu\nu} = \frac{1}{2}
[ (  \frac{\partial U_{\mu}(\vec{r})}{\partial x_{\nu} } +
\frac{\partial U_{\nu}(\vec{r})}{\partial x_{\mu}}  )
+ \sum_{\rho} \frac{\partial U_{\rho}} {\partial x_{\mu}}
\frac{\partial U_{\rho}} {\partial x_{\nu}} ]
\eqno(B1)
$$
and we retain the `geometric
nonlinearity', showing when it is negligible later.
The symmetry-adapted  compressional ($\phi_1$), deviatoric
($\phi_2$) and shear ($\phi_3$) strains are
$$\phi_{1}/c_1 = \frac{1}{2}(\phi_{xx}+\phi_{yy}), \
\phi_{2}/c_2 =\frac{1}{2}(\phi_{xx} -\phi_{yy});$$
$$\phi_{3}/c_3=\frac{1}{2}(\phi_{xy} + \phi_{yx}), \eqno(B2)$$

\noindent where $c_1,c_2,c_3$ are symmetry-specific constants.
The free energy  depends on the order-parameter through a Landau term,
and has non-$OP$ and gradient contributions:

$$F = F_{Landau} + F_{non}+ F_{Grad}. \eqno(B3)$$

The order parameter Landau energy is

$$F_{Landau} = \int d^{3}r [ \sum_{\ell}   \frac{1}{2}
B^{(2)}(T) \phi_{\ell}^{2}   + F_{poly}  ],  \eqno(B4a)$$

\noindent where $B^{(2)}(T)=B_0 (T-T_c)$ vanishes at a characteristic
temperature
$T_c$, and the $F_{poly}$ is a temperature independent $OP$ polynomial with 
powers higher than
quadratic, that is different for the $SR$ and $TR$ cases.
The non-$OP$ contribution is
$$F_{non } = \frac {1}{2}  \int d^{3}r \sum_i B_i \phi_{i}^{2}
\eqno(B4b))$$
while the gradient term is

$$F_{Grad} = \int d^{3}r  \sum_{\ell}   \frac{1}{2}
K [\vec{\nabla} {\phi_{\ell}}]^{2} .   \eqno(B4c)$$

We assume uniformity in the $z$-direction, of thickness $h$,
and pass over to a 2D
reference lattice of lattice constant $a_o$ (square or triangular for $SR$
and $TR$ cases), thus unit-cell strains are tensorial  variables on the dual
lattice. Derivatives are converted to discrete lattice differences, and
displacements are scaled in $a_{o}$. Thus

$$\frac{\partial}{\partial x_{\mu}} \rightarrow
\frac{1}{a_o}\Delta_{\mu};
\int d^{3}r \rightarrow h a_o^{2}\sum_{\vec{r}} .
\eqno(B5)$$
We  further scale strains in a typical value
$\lambda$ chosen for convenience later, so
a scaled strain  tensor $E_{\mu\nu}$ is then defined through

$$\phi_{\mu\nu} = \lambda E_{\mu\nu} =
\frac{\lambda}{2} [
\{\Delta_{\mu}u_{\nu} + \Delta_{\nu}u_{\mu}\} +
\lambda \sum_{\rho} \Delta_{\mu} u_{\rho} \Delta_{\nu}u_{\rho} ],
\eqno(B6)$$
so
$\nabla_{\mu} U_{\nu} \rightarrow \lambda \Delta_{\mu} u_{\nu}$ where
$\vec{u}$ is a dimensionless scaled displacement.
The scaled $OP$ and non-$OP$ strains are
$$\phi_{\ell} =\lambda \varepsilon_{\ell}, ~~\phi_i = \lambda
e_i. \eqno(B7)$$

The free energy (B3) can be written in terms of these scaled strains as

$$F = F_0 (\{ \varepsilon_{\ell} \}, \tau) + f(\{ e_i \} ) +
F_{grad} (\{ \vec{\Delta} \varepsilon_{\ell} \} ), \eqno(B8)$$
where
$$F_{non}/E_{0} \equiv f =  \frac{1}{2} \sum_{\vec{r},i} a_{i}
e_{i}^{2}, \eqno(B9a)$$
$$F_{Grad}/ E_0 \equiv F_{grad} =  \sum_{\vec{r},\ell}
\frac{K_0}{2} (\vec{\Delta} \varepsilon_{\ell})^2. \eqno(B9b)$$

\noindent Here we multiply and divide by an energy density $D_0$ chosen
later, to get the overall energy scale $E_0 = h {a_0}^2 D_0$ (this drops
out in the end).  The scaled parameters are
$$ a_i =  \frac{B_{i}\lambda^2} {D_0};
~~K_0 =\frac{ K \lambda^{2} }{(a_{o}^{2} D_0)}. \eqno(B10)$$
The scaled $OP$ free energy is
$$F_{Landau}/ E_0 \equiv F_0 = \sum_{\vec{r},\ell} (\tau -1)\varepsilon_{\ell}^{2}
+ F_{00}, \eqno(B11a)$$
where  $F_{00} \equiv F_{poly}/E_0 +  \sum_{\vec{r},\ell}
{\varepsilon_{\ell}}^2$. The scaled
temperature $\tau$
contains a physical temperature $T_0$ that fixes where $\tau =1$:
$$\tau \equiv \frac{(T- T_c)}{(T_0 - T_c)}
, ~~(\frac{ B_0 \lambda^{2} }{D_0} ) \equiv \frac{1}{(T_0 - T_c )} . 
\eqno(B11b)$$
We now consider the $SR$ and $TR$ symmetries separately.
\\
\noindent{\it SR case scaling:}

For the $SR$ case, $N_{op}=1$,
there is only the deviatoric strain, as $\phi_2$ as the order parameter
and the polynomial is
$$F_{poly} = \int d^3 r [
-B^{(4)}\phi_{2}^{4} + B^{(6)}{\phi_2} ^{6} ].
\eqno(B12)$$
In scaled form, the term in (B11a) is

$$F_{00} = \sum_{\vec{r}} [ {\varepsilon_2}^2 -C_0 {\varepsilon_2}^4 +
{\varepsilon_2}^6 ], \eqno(B13)$$

\noindent where we have factored out an energy density
$D_0=B^{(6)}{\lambda}^6$ so the coefficient of the sixth order term is
unity, and $C_0 \equiv B^{(4)} \lambda^{4}/D_0$. Now we
choose $\lambda$
to fix $C_0$ so that for three ($\alpha=+,-,0$) roots
$\bar{\varepsilon_{2}}^\alpha$  at $\tau = 1$, the conditions
of degeneracy
$$ F_{00} (\{ \bar{\varepsilon}^{(\alpha)}_{\ell} (1)\}) =
F_0 (\{ \bar{\varepsilon}^{(\alpha)}_{\ell} (1)\}, \tau=1)=0
 , \eqno(B14)$$ 
and normalization
$$\sum_{\ell} [\bar{\varepsilon}^{(\pm)}_{\ell}(1)]^2=1 \eqno(B15)  $$
 
\noindent are satisfied, and hence determine the
typical strain $\lambda$ and all scaled parameters. For the $SR$ case,

$$ C_0 =2 ; \ \lambda = (B^{(4)}/ 2 B^{(6)})^{\frac{1}{2}}, \
D_0 = (B^{(4)}/2)^{3}/ (B^{(6)})^2,   \eqno(B16)$$

\noindent with the $C_0$ `completing the square' in $F_{00}$ and
thus the $OP$ free energy is

$$F_0= \sum_{r}(\tau - 1)\varepsilon_{2}^{2} + \varepsilon_{2}^{2}(
\varepsilon_{2}^{2} - 1)^{2}.  \eqno(B17)$$

\noindent The $\tau = 1$ roots, $\bar{\varepsilon}^{(\pm)}_2 = \pm 1$,
then manifestly satisfy the conditions (B14) and (B15).

To get an idea of parameters, we use FePd shape memory alloy values
\cite{Kartha}, for the $SR$ case with energy densities in units of
$ergs/cm^3$,
$B_{1}=1.4 \times 10^{12}$,
$B_{2}=2.8 \times 10^{12}$,
$C_{2}=1.7 \times 10^{12}$,
$D_{2}=3 \times 10^{17} $,
$\frac{K}{a_{o}^{2} } = 2.5 \times 10^{11}$, and
$B_{o} = 2.4 \times 10^{9} \frac{ergs}{cm^{3}\deg K}$. This
gives, from (B6), the typical strain value
$\lambda = 0.02$; elastic constants $a_{1} =155 \approx a_3 /2$;
$OP$ variation scale $\sqrt{K_0}\approx 5$; an elastic energy density
$\frac{E_0}{a_{0}^2h}=D_{0}=3.8\times 10^{6} \frac{ergs}{cm^{3}}$,
and a temperature separation $T_{0} -T_{c}=7$ Kelvin. (The magnitude of 
$D_0$ corresponds to a magnetic energy density $\frac{H^2}{8\pi}$ for fields
$\sim 1$ Tesla.) 
Note that $\lambda << 1$, thus in (B6) we may drop the
`geometric nonlinearity' and work with the linear Cauchy strain
tensor, as  in the text, that satisfies the simple
St. Venant compatibility condition. External stresses (eg. compressional)
enter as $F \rightarrow F + \sum_{r} p_{1}e_{1}$ with scaled pressure
$p_1=1$ corresponding to $\frac{D_0}{\lambda}=0.02 GPa$.
\\ 

\noindent{\it Inertial and Damping terms:}

Using the same kind of transformations, the dimensionless inertial $( T)$
and damping $( R^{tot})$ contributions to the Lagrangian  are:

$${\it T }= \frac{1}{2} \int d^{3}r \rho_{m} ( \frac{\partial U}{\partial
t})^{2}/
E_{o} = \frac{1}{2}\sum_{\vec{r}}\rho_0 \dot{u}^{2} \eqno(B18a)$$
and

$${R^{tot}} =\frac{1}{2}\int d^{3}r\sum_{j=1,2,3}B_{j}(\frac{\partial\phi_j}{\partial t})^2/E_0 
$$

$$ \ \ \ = \frac{1}{2} \sum_{r} [a'_{1}\dot{e}_{1} ^{2} +
a'_{3}\dot{e}_{3} ^{2} +
A'_{2}\dot{\varepsilon}_{2}^{2}], \eqno(B18b)$$

\noindent where the dots are dimensionless time derivatives, and
we introduce a characteristic time unit $t_o$, with
dimensionless density $\rho_0$ and friction coefficients $a'_{1,3}$
and $A'_{2}$ defined as

$$\rho_0 = \frac{\rho_{m}\lambda^{2}a_{o}^{2} }{t_{o}^{2}D_{o}},
~~\ \frac{a'_{1,3}}{a_{1,3}}=\frac {B'_{1,3} }{B_{1,3}t_{o}^2}, \
~~A'_{2}=\frac {B'_{2}/B_{o}t_{o}^{2}}{(T_{o}-T_c)}.  \eqno(B19) $$
Here we have used (B10), (B11b) for $a_{1,3}$ and $(T_{o}-T_c)$.
Wave propagation crossing a nanometer in a picosecond
corresponds to a sound speed of $1000$ m/sec.
We take $t_o$ to be of the order of inverse  phonon
frequencies
and the scaled friction coefficients $a'_{1,3}$, $A'_{2}$ to then be
less than unity.  With a mass density of $\rho_{m}=10$ $gm/cm^3$, lattice
constant $a_{o}\sim 3$ $\mbox{\AA}$ and $t_{o}\sim 10^{-12}sec$  the
dimensionless density, or ratio of kinetic and elastic energy densities is
$\rho_0 \sim 1$.
We will work with parameters $\rho_0 = 1, K_0= 1, t_0 = 1$ picosecond,  a 
fixed ratio $a_3/a_1 = 2.1$, and $a_1 = 100$ or $10$, with  $a'_1,
A'_2,a'_3$ as unity or less. \\

\noindent{\it $TR$ case scaling:}

For the $TR$ case we start from \cite{Reid,JacobsHO} a free energy density
in $F_{poly} \sim-B^{(3)}(\phi_{2}^3 - 3\phi_{2}\phi_{3}^2
) +B^{(4)} (\phi_{2}^{2} + \phi_{3}^{2})^4$.
We follow the same  procedure as above : (i) scaling strains as in (B7);
(ii) pulling out a common factor $D_0 \equiv B^{(4)} \lambda^4$;
(iii) choosing $\lambda$ in  $C_0 \equiv B^{(3)} \lambda^3 / D_0$ to
satisfy (B15).
This gives
$$ C_0 = 2, \  \lambda = B^{(3)}/2 B^{(4)}, \  D_0 = (B^{(3)}/2)^4 /
(B^{(4)})^3. \eqno(B20)$$

The scaled free energy is then

$$F_{00} = \sum_{\vec{r}} ({\varepsilon_2}^2 + {\varepsilon_3}^2)
- 2 (\varepsilon_{2}^{3} - 3 \varepsilon_{2}\varepsilon_{3}^2) +
({\varepsilon_2}^2 + {\varepsilon_3}^2)^2
\eqno(B21a) $$
$$ = \sum_{\vec{r}} 3(1 + 2 \varepsilon_2) [ {\varepsilon_3}^2
- \frac{1}{3} (1 - {\varepsilon_2}^2)^2 ] + [{\varepsilon_2}^2
+ {\varepsilon_3}^2  -1]^2.  \eqno(B21b)$$

\noindent Here the second form explicitly displays the (B14),
(B15) conditions,
for $\tau = 1$ , when the roots are $({\varepsilon_2}^{(0)}
,{\varepsilon_3}^{(0)}) = (1,0)$;  $({\varepsilon_2}^{(\pm)}
,{\varepsilon_3}^{(\pm)}) = (-\frac{1}{2}, \pm
\frac{\sqrt{3}}{2})$, and lie on a unit circle.
\\ 

\noindent{\it Connection to other scalings:}

The $SR$ free energy of Ref. \onlinecite{JacobsSR} can be written as our 
(B9),
(B17), (B18b) by a scaling of strains in $\alpha = 10^6$, giving an
overall
factor $\frac{1}{2} \alpha^2$ absorbed in the $TDGL$ time.
(The elastic and frictional constants are then half our $a_i,a'_i$.)
A similiar $TR$ scaling in $\alpha = 10^3$ yields our (B21), with a
common factor of $\frac{1}{2}$ relating elastic/friction constants
\cite{JacobsHO}.  Similar scaling can be performed for other symmetries.
\\\

\section{TRUNCATED DISPLACEMENT DYNAMICS AND TDGL EQUATIONS}

We have shown that the $BG$ equations, dropping strain-accelerations in an 
$\omega-k$ regime, yield $TDGL$ equations. On the other hand, dropping 
{\it displacement}-accelerations in (2.8) or A3 yield equations stated 
to be different \cite{JacobsHO,JacobsSR} from the $TDGL$ form. 
In this Appendix, we 
demonstrate their equivalence to $TDGL$. 

More generally, the truncation is like  dropping  all the inertial
terms in the
Lagrange-Rayleigh equations (3.1), (3.2):
$$\sum_{\nu}
\partial_\nu\sigma_{\mu\nu}=-\sum_{\nu}\partial_\nu\sigma'_{\mu\nu}
\eqno(C1)$$

\noindent given as a balance \cite{JacobsSR} between derivatives of the
stress tensor
$\sigma_{\mu\nu}
= \delta L/\delta E_{\mu\nu}$
and the damping force tensor
$\sigma'_{\mu\nu}=\delta R^{tot}\delta \dot{E}_{\mu\nu}$.
Clearly \cite{JacobsSR}, one cannot proceed in 2D by simply dropping the
$\partial_{\nu}$ derivatives in (C1).
Such a procedure would give $\sigma_{\mu\nu}=-\sigma'_{\mu\nu}$, which
is not quite correct. For the
$SR$ transformation, for example, this is $\dot{e}_{1,3} =
-({a_{1,3}}/{a'_{1,3}}) e_{1,3}$; and
$\dot{\varepsilon_{2}}=-\frac{1}{A'_2}\frac{\partial
F}{\partial\varepsilon_2}$,
where $F \sim \varepsilon_{2}^{6}$ is just the local triple-well
free energy. This is the kind of overdamped dynamics \cite{Chen}, without
compatibility contributions, that would
emerge if all the strains were independent.

We now show that the truncated displacement dynamics (C1) is
in fact a $TDGL$ dynamics.\\

\noindent{\it (a) $TR$ case displacement truncations:}
Dropping $\{\ddot{u_{\mu}}\}$ and keeping $\Delta_{\mu}\dot{u}_{\nu}$
in (2.8) yields \cite{JacobsHO} equations that

\noindent in Fourier space are 
$$
a_{1}'k_x\dot{e}_1+A_{2}'k_x\dot{\varepsilon}_2+A_{3}'k_y\dot{\varepsilon}_3=
-\{ a_1k_xe_1+k_xF_2+k_yF_3 \}, \eqno(C2a)
$$
$$
a_{1}'k_y\dot{e}_1-A_{2}'k_y\dot{\varepsilon}_2+A_{3}'k_x\dot{\varepsilon}_3=
-\{ a_1 k_ye_1-k_y F_2+k_x F_3 \}. \eqno(C2b)
$$
Using  compatibility (3.18d) we can  eliminate
$$e_1=-[Q_2(\vec{k})\varepsilon_2+ Q_3(\vec{k})\varepsilon_3]/Q_1(\vec{k}),
\eqno(C3)$$
where
$Q_2/Q_1 =(\frac{k_x^{2}-k_y^{2}}{k^2});
Q_3/Q_1 =\frac{2k_xk_y}{k^2}$. Further,
(C2) can be written in matrix form as
\[
\b{\b{M}}
\left(
\begin{array}{cc}
\dot{\varepsilon_2} \\
\dot{\varepsilon_3}
\end{array}
\right)
=
-\left(
\begin{array}{cc}
k_x & k_y \\
k_y & -k_x
\end{array}\right)
\left(
\begin{array}{cc}
F_2+F_2^{c} \\
F_3+F_3^{c}
\end{array}\right),   \ \ \ \ \ \ \ \ \ \ \ \ \ \ (C4)
\]
where $M_{\ell \ell'}$ is defined as $M_{22} = k_x [A'_2 - a'_1 Q_2 /Q_1]$;
$M_{33} = -k_x [A'_3 -(k_y / k_x) a'_1 (Q_3/Q_1)]$; $M_{23}=
k_y [A'_3 - a'_1 (k_x /k_y) (Q_3 / Q_1)]$; $M_{32}= k_y [A'_2 +a'_1
(Q_2 /Q_1)]$. Here
$F^{c}_{2},F^{c}_{3}$
are chosen to match the $RHS$ terms of (C2), so that
$F_2^{c}=a_{1}[(Q_{2}^2/{Q_1}^2)\varepsilon_2 + (Q_{2}Q_{3}/{Q_1}^2)
\varepsilon_3],$ and $F_{3}^{c}=a_1 [(Q_{2}Q_{3}/{Q_1}^2)\varepsilon_2
+a_{1}(Q_{3}/Q_1)^{2}\varepsilon_3] $
and can be written as derivatives of a compatibility potential $F^{c}$:
$$F_{2,3}+F_{2,3}^{c}=\frac{\partial
(F+F^{c})}{\partial{\varepsilon_{2,3}}^{\star}
(\vec{k})}, \eqno(C5)$$
where
$F^{c}={1\over 2}a_1\sum_{\vec{k}}
{U^c}_{\ell \ell'}
\varepsilon_{\ell}(\vec{k},t){\varepsilon^*}_{\ell'}(\vec{k},t)$
with ${U^c}_{\ell \ell'} \equiv Q_{\ell} Q_{\ell'}/{Q^2}_1$.
Inverting the matrix $\b{\b{M}}$ of (C4) yields (5.2), and then the $TDGL$
link is as in the text. For a choice \cite{JacobsHO} $a'_1 =0$,
one gets the ordinary (local, instantaneous) $TDGL$ equation,
namely (5.5). \\

\noindent{\it SR case displacement truncations:}

\begin{figure}[h]
\epsfxsize=7cm
\epsffile{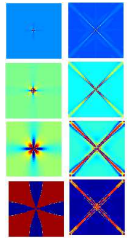}
\caption
{
Evolution from single-site initial condition for $SR$ case under
ordinary $TDGL$
dynamics. Times are $t=0.13,0.16,0.18,0.24$ picosecs, with
$\Delta t=10^{-4}$. The
initial condition is $\varepsilon_2 (\vec{r}, t=0) = 0.0001$ at a single
site and zero elsewhere. Non-$OP$ friction constants are $a'_1 =0= a'_3$.
Left column: `Soft' case, $a_1 = 2, a_3=2, \tau = -50,
A'_2 =2$. Right column: `Hard' case,   $a_1 = 2000, a_3 = 2000, \tau = -50, 
A'_2 =2$. (Since
colors are relative, the background changes, with
changes in evolving average intensity.)
}
\label{fig9}
\end{figure}

Similiarly for the $SR$ case, dropping displacement accelerations 
compared to gradients of displacement velocities
\cite{JacobsSR} in (A3) yields the Fourier space equation, 
with $a_{i \omega} \equiv a_1 - i \omega a'_i$,

$$(\frac{c_{1}}{c_2})a_{1\omega}e_1 +
(\frac{c_{3}}{c_2})a_{3\omega}(\frac{k_y}{k_x})e_{3} = - [ -i\omega A_{2}'
+
F_2 ], \eqno(C7a)$$

$$(\frac{c_{1}}{c_2})a_{1\omega}e_1 +
(\frac{c_{3}}{c_2})a_{3\omega}(\frac{k_x}{k_y})e_{3} =  [ -i\omega A_{2}'
+
F_2 ]. \eqno(C7b)$$

\noindent Then compatibility $e_{1} + (\frac{Q_2}{Q_1})\varepsilon_{2}
+ (\frac{Q_3}{Q_1})e_3 = 0$ gives

$$e_{1} = (\frac{Q_{1}}{Q_{3}})
{(\frac{a_{3\omega}}{a_{1\omega}})} e_3;$$
$$ \\ e_{3} =
-[\frac{(\frac{Q_{2}}{Q_3})}
{1 + (\frac{a_{3\omega}}{a_{1\omega}})(\frac{Q_{1}}{Q_{3}})^2 }]
\varepsilon_2. \eqno(C8)$$

\noindent Hence

$$ -i \omega A'_2 \varepsilon_2 = - F_2 + \left[
\frac{a_{3\omega}(\frac{Q_2}{Q_3})^2}{1 + (\frac{a_{3\omega}}{a_{1\omega}})
(\frac{Q_{1}}{Q_{3}})^2 }\right] \varepsilon_2.  \eqno(C9)$$

\noindent Separating real and imaginary parts of the square brackets, and 
using
the
notation of (3.35),
$$ -i\omega \varepsilon_{2} = - \lambda (\hat{k},\omega)[F_{2} +
a_1{ U}^{c}(\hat{k},\omega^2,0)\varepsilon_{2}],  \eqno(C10))$$
where now $\lambda (\hat{k},\omega) = 1/[ A'_2 + a'_1 \eta^c (\hat{k},
\omega^2, 0)]$.
This is manifestly a
generalized $TDGL$ as in (5.6a), but with a partial truncation of the
kernel, that drops the (resonant) inertial delay terms
$(\rho_0/a_3)(\omega/k)^2$ and keeps
only frictional retardation.
It is thus valid for a narrowly restricted, intermediate length and time
regime roughly estimated as $L_D (t) > L_p (t) > L$ and $t < t_f$ (see Sec.V). 
Taking a more well defined limit of late times (dropping all $\omega^2$) 
the static kernel $U^c (\hat{k},0,0)$
becomes a good approximation, and the nonlocal $TDGL$ (5.4) with 
$\ell,\ell'=2$ holds.
For $a'_{1,3}/a_{1,3} << A'_2$ as for `hard' systems \cite{JacobsSR} this
collapses to a
{\it local} $TDGL$, as in (5.8).
Thus overdamped displacement equations
\cite{JacobsHO,JacobsSR} are $TDGL$ 
equations in disguise and not a new dynamics. As expected, simulation of
$TDGL$ equations produce the same textures \cite{JacobsHO,JacobsSR}
from similar initial conditions.

Figure 9 shows ordinary $TDGL$ simulations with an initial condition
\cite{JacobsSR} of
strain nonzero at a single point and zero elsewhere.
This is like a stress applied at a single point and then removed.
Flower-like or diagonal-cross textures similiar to
Ref.\onlinecite{JacobsSR} are obtained, for both `soft' and `hard' materials.

Reference \onlinecite{JacobsSR} could not reproduce
a $TDGL$ structure given in Fig. 4b of Ref.  
\onlinecite{ourRC}, part of a multi-panel figure that displays stress 
effects
for soft materials. In the
intermediate-temperature ($1 >\tau > 0$) phase,
Ref. \onlinecite{ourRC} considered  
the effect of two Lorentzian
deviatoric stresses, fixed and continuously maintained. 
The motivation was to see if a stress seed analogous to a seed crystal
in a supercooled melt, could yield stress-induced martensitic
twins even for positive $\tau=0.3$. The Ref. \onlinecite{JacobsSR}
simulations did not have the applied constant stress of 
Ref. \onlinecite{ourRC}, and
moreover, were at very low temperatures $\tau=-50$.
We conclude that the difference in results is due
to a difference in states investigated, and {\it not} a difference in
dynamics, which is $TDGL$-like in both cases.


\begin{references}

\bibitem{wadhawan} V. K. Wadhawan, {\it Introduction to Ferroic
Materials} (Gordon and Breach, Amsterdam, 2000).

\bibitem{martens} E.K.H. Salje, {\it Phase Transformations in Ferroelastic
and Co-elastic Solids} (Cambridge University Press, Cambridge, U.K., 1990).

\bibitem{nishi} Z. Nishiyama, {\it Martensitic Transformations}
(Academic, New York, 1978).

\bibitem{otsuka} {\it Shape Memory Materials}, edited by K. Otsuka
and C. M. Wayman (Cambridge University Press, Cambridge, U.K., 1998).

\bibitem{BG} G.S. Bales and R.J. Gooding, Phys. Rev. Lett. {\bf 67}, 3412
(1991).

\bibitem{Reid} A.C.E. Reid and R.J. Gooding, Phys. Rev. B {\bf 50},
3588 (1994); Physica A {\bf 239}, 1 (1997);
B.P. Van Zyl and R.J. Gooding, Metall. Mat. Trans. A
{\bf 27}, 1203 (1996).

\bibitem{Chen} S.K. Chan, J. Chem. Phys. {\bf 67}, 5755 (1977).

\bibitem{Khatch}  Y. Wang and A.G. Khachaturyan, Acta Mater. {\bf 45},
759 (1997); S. Semenovskaya, Y.M. Zhu, M. Suenaga, and A.G. Khachaturyan,
Phys. Rev. B {\bf 47}, 12182 (1993); Y. Wang and A.G. Khachaturyan, Acta
Mater. {\bf 45}, 759 (1997).

\bibitem{kerr} W.C. Kerr, M.G. Killough, A. Saxena, P.J. Swart,
and A.R. Bishop, Phase Trans. {\bf 69}, 247 (1999).

\bibitem{ourRC}  S.R. Shenoy, T. Lookman, A. Saxena, and
A.R. Bishop, Phys. Rev. B {\bf 60}, R12537 (1999).

\bibitem{our3DPRL} K. \O. Rasmussen, T. Lookman, A. Saxena,
A.R. Bishop, R.C. Albers, and S.R. Shenoy, Phys. Rev. Lett.
{\bf 87}, 055704 (2001).  The $N_c=6$ compatibility equations are 
actually in two sets of 3 equations (see L.E. Malvern in Ref. 29).

\bibitem{Yama} Y. Yamazaki, J. Phys. Soc. Jpn. {\bf 67}, 1587
(1998); {\bf 67}, 2970 (1998).

\bibitem{Onuki} A. Onuki, J. Phys. Soc. Jpn. {\bf 68}, 5 (1999).

\bibitem{Anantha} R. Ahluwalia and G. Ananthakrishna, Phys. Rev. Lett.
{\bf 86}, 4076 (2001).

\bibitem{Madan} M. Rao and S. Sengupta, Phys. Rev. Lett. {\bf 78}, 2168
(1997); M. Rao and S. Sengupta, cond-mat/0107098.

\bibitem{JacobsHO} S.H. Curnoe and A.E. Jacobs, Phys. Rev. B {\bf 63},
09410 (2001).

\bibitem{JacobsSR} S.H. Curnoe and A.E. Jacobs, Phys. Rev. B {\bf 64},
064101 (2001).

\bibitem{YBCO} J.A. Krumhansl, in {\it Lattice Effects in High-$T_c$
Superconductors}, eds. Y. Bar-Yam, T. Egami, J. Mustre de Leon, and
A.R. Bishop (World Scientific, Singapore, 1992).

\bibitem{asamitsu} A. Asamitsu, Y. Moritomo, Y. Tomioka, Y. Arima,
and Y. Tokura, Nature (London) {\bf 373}, 407 (1995).

\bibitem{moore} A. Moore, J. Graham, G.K. Williamson, and G.R. Raynor,
Acta. Metall., {\bf 3}, 579 (1955).

\bibitem{bhk}
G. R. Barsch, B. Horovitz, and J. A. Krumhansl, Phys. Rev. Lett. {\bf
59}, 1251 (1987); B. Horovitz, G.R. Barsch and J.A. Krumhansl, Phys. Rev.
B {\bf 43}, 1021 (1991).

\bibitem{sugiyama}
M. Sugiyama, Ph.D. Thesis, Osaka University, 1985;
S. Muto, R. Oshima, and F.E. Fujita, Acta Metall. Mater. {\bf 38},
685 (1990).

\bibitem{sb} A. Saxena and G.R. Barsch, Physica D {\bf 66}, 195 (1993).

\bibitem{Kartha} S. Kartha, J.A. Krumhansl, J.P. Sethna, and L.K.
Wickham, Phys. Rev. B {\bf 52}, 803 (1995).

\bibitem{planes} L. Carrillo, L. Ma\~nosa, J. Ortin, A. Planes, and
E. Vives, Phys. Rev. Lett. {\bf 81}, 1889 (1998).

\bibitem{LL} L.D. Landau and E.M. Lifshitz, {\it Theory of Elasticity}
(Pergamon, Oxford 1982).

\bibitem{eshelby} J. D. Eshelby, Solid State Phys. {\bf 3}, 79
(1957).

\bibitem{venant} C.L.M.H. Navier, \emph{R\'{e}sum\'{e} des
Le\c{c}ons sur l'Application de la M\'{e}canique}, 3\`{e}me edition
avec des notes et des appendices par A.J.C. Barr\'{e} de
Saint-Venant (Dunod, Paris, 1864).
\bibitem{compat} D.S. Chandrasekharaiah and L. Debnath,
{\it Continuum Mechanics} (Academic, San Diego, 1996) p. 218;
S.Timoshenko, {\it History of Strength of Materials} (McGraw-Hill,
New York, 1953) p. 229;  E.A.H. Love, {\it A Treatise on the
Mathematical Theory of Elasticity} (Dover, New York, 1944);
L.E. Malvern, {\it Introduction to the Mechanics of a Continuous
Medium} (Prentice Hall, New Jersey, 1969); I. S.  Sokolnikoff,
Mathematical Theory of Elasticity (McGraw-Hill, New York, 1946).

\bibitem{Baus}  M. Baus and R. Lovett, Phys. Rev. Lett. {\bf 65},
1781 (1990); Phys. Rev. Lett. {\bf 67}, 406 (1991); Phys. Rev. A
{\bf 44}, 1211 (1991).

\bibitem{lattice} 
In Fourier space, with a unit lattice constant, the derivatives are
replaced as
$\Delta_{x,y}^2 \rightarrow 2 - 2 \cos k_{x,y} = 4 \sin^2 \frac{k_{x,y}}{2}$.
This Brillouin-zone Fourier substitution is automatically taken care
of in simulations by FFT subroutines. To save on notation, we will
write in the text ${\Delta_{x,y}}^2 \rightarrow {k_{x,y}}^2$ with the
understanding that the long-wavelength limit ${k_{x,y}}^2$ actually stands
for the
sinusoidal lattice function $ 4 \sin^2 \frac{k_{x,y}}{2}$. Thus, for example,
the discrete Laplacian will be written as ${\vec{\Delta}}^2 \rightarrow 
{\vec{k}}^2$ instead of $4 - 2\cos k_{x} - 2 \cos k_{y} = 4 \sin^{2} 
\frac{k_{x}}{2} +4 \sin^{2} \frac{k_{y}}{2}$. 
{\it Functions} like $Q_s(\vec{k})$, $M(k)$ will also use this compact 
notation.  Of course, {\it labels} of points in the Brillouin zone $\vec{k}$, 
still stand for the actual wave vector $\vec{k}=|\vec{k}|\hat{k}$.
To emphasize that textural
nonuniformity results from the large-scale compatibility properties of
the (`anisotropic continuum') solid, rather than from trivial
small-scale discreteness, we use the long wavelength $U(\hat{k})$
in simulations.
 
 
\bibitem{Fscaling} G.R. Barsch and J.A. Krumhansl, Metallurg.
Trans. A {\bf 19}, 761 (1988).


\bibitem{bc}
For periodic boundary conditions, the strains in a box
${L_0}\times {L_0}$ have the same constant
values at opposite box sides: $\varepsilon_\ell (x,y=0, t) =
\varepsilon_\ell (x,y=L_0, t);
\varepsilon_\ell (x=0,y, t)=\varepsilon_\ell (x=L_0,y, t)$.
This is satisfied by expanding in a complete set
of plane wave states, as is natural for
simulations using FFT routines: $e^{i k_x L_0} = e^{i k_y L_0} = 1$ fixes
the wavevector spectrum. The second derivative of strains is
consequently periodic.
Net dissipation must only occur from Rayleigh terms inside the box,
and not over the boundaries \cite{Reid}. The periodic boundary
conditions enforce this requirement. The $BG$
equation is also a continuity-type equation, with
$\rho_0 \dot{\varepsilon}_{\ell} (\vec{r},t)\equiv n (\vec{r},t)$ as a
`momentum density'
and $\frac{1}{4}{c_{\ell}^{2}}\vec{\Delta}[ (\partial (F + F^c)/{\partial
\varepsilon_{\ell} } +
\partial (R + R^c)/{\partial \dot{\varepsilon}_{\ell}
})]\equiv \vec{J}(\vec{r},t)$ as a `momentum current'. (Note
the
$\partial F/{\partial \varepsilon_{\ell} }$ term contains a
${\Delta}^2 \varepsilon_{\ell}$ contribution.)
The generalized $BG$ equation is then $\dot{n} = \vec{\Delta}\cdot\vec{J}$. 
Integrating over the box yields normal components of $\vec{J}$
that add up to zero: the box surfaces have no net frictional losses.
(Since periodic boundary conditions are foldings of opposite
edges onto each other, there are no real boundaries.) 
The required condition \cite{Reid} $d(T+V)/dt = -2R$, that
energy loss is determined only by the bulk Rayleigh dissipation rate,
follows at once, using the strain-acceleration expression (4.4a) of
$BG$ dynamics.


\bibitem{surface} 
For finite systems the no-defect compatibility condition 
holds, both in the bulk and at the surfaces. Periodic boundary
conditions on strains reveal the bulk compatibility potential
from internal non-OP strains. The surfaces induce extra non-OP
strains that can be shown to produce additional surface compatibility
potentials \cite{ourRC}, that decrease with increasing system size,
and can also
be written in terms of the bulk OP strains. These will be considered
elsewhere, and we focus on periodic boundary conditions, and bulk
compatibility, for simplicity.  

\bibitem{Hatch}
D.M. Hatch, T. Lookman, A. Saxena and S.R. Shenoy, preprint (2002).


\bibitem{U(R)} The Fourier integral is performed by rotating the wavevector
axes by $45$ degrees, defining $\tan \gamma \equiv a_1 /a_3$ and doing
first the $k_y$ and then the $k_x$ contour integrals, and taking the
relevant real parts of the residues.  We thank Yuri B. Gaididei (private
communication) for help with this derivation.

\bibitem{retard}
J.D. Jackson, {\it Classical Electrodynamics} (Wiley, New York, 1999).


\bibitem{Delay}
Time delays are also contained in the $TR$ case dynamics, but with
mutually-nonlinear couplings in both $OP$ strain components there could
be a spectrum of mutually-driving frequencies.



\bibitem{simul} We use ${L_0 \times L_0 = 128}^2$ lattices with
periodic boundary conditions on 
the strain \cite{bc} and Euler routines, with step sizes and parameters as in
the figure captions. With notional time units (Appendix B) as
picoseconds and number of steps $N_t$ we  go up to
$t \equiv \Delta t N_t \sim 1000$ picoseconds in some cases. Diagnostics
for monitoring
textural evolution
included the free energy per unit cell $E \equiv F (t)/{L_0}^2$; forces
$\partial F/\partial \varepsilon_\ell (t)$; spatially
averaged strains $\langle \varepsilon_\ell (\vec{r},t) \rangle$,
$\langle e_i (\vec{r},t) \rangle$; and
max-min values  of $\{\varepsilon_\ell (\vec{r},t)\}$, $\{
e_i (\vec{r},t)\}$. Unless specified, the initial conditions were randomly
varying strains with zero spatial average, and amplitudes less than
$\sim 0.2$. When non-OP strains are obtained dynamically, at finite times, 
we use that average in the diagnostics.

\bibitem{LANL} A. Saxena, T. Lookman, A.R. Bishop, and S.R. Shenoy,
Los Alamos National Laboratory publication LA-UR-99-336 (May 1999).


\bibitem{ourPhysica} A. Saxena, T. Lookman, S.R. Shenoy, and A.R. Bishop,
Mater. Sci. Forum {\bf 327-328}, 385 (2000).



\bibitem{Wen}
Y.H. Wen, Y.Z. Wang, and L. Q. Chen, Philos. Mag. A {\bf 80}, 1967 (2000).

\bibitem{ourAbstract}
T. Lookman, A. Saxena, S. R. Shenoy, and D. M. Hatch,
Session T8 abstracts, Materials Research Society (MRS) Fall
Meeting, Boston, Nov. 2001.


\bibitem{manolikas}
C. Manolikas and S. Amelinckx, Phys. Stat. Sol. (a) {\bf 60}, 607 (1980);
{ibid}. {\bf 61}, 179 (1980).

\bibitem{Anom} G.R. Barsch and J.A. Krumhansl, Proc. ICOMAT-92, eds.
C.M. Wayman and J. Perkins (Monterey Institute of Advanced
Studies, CA, 1993) p.53; M. Sato, B.H. Grier, S.M. Shapiro,
and H. Miyajima, J. Phys. F.: {\bf 12}, 2117 (1982);
T.R. Finlayson, M. Mostoller, W. Reichardt, and H.G. Smith,
Solid State Commun. 53, 461 (1985).

\bibitem{FP} H. Risken, {\it The Fokker-Planck Equation: methods of
solution and applications} (Springer-Verlag, Berlin 1989);
H. Haken, {\it Synergetics: an introduction} (Springer-Verlag, Berlin,
1983).

\bibitem{Bala} V. Balakrishnan, Pramana (Bangalore) {\bf 11}, 379 (1978);
{\bf 13}, 547 (1979); V. Balakrishnan and C.E. Bottani eds., {\it Mechanical
Properties of Solids: Plastic Instabilities}
(World Scientific, Singapore, 1986).

\bibitem{osc} H. Goldstein, {\it Classical Mechanics} (Addison-Wesley,
Reading MA 1980).

\bibitem{HH}  P.C. Hohenberg and B. Halperin, Rev. Mod. Phys.
{\bf 49}, 435 (1977).

\bibitem{complexosc} Each of the ${L_0}^2$ sites in the Brillouin
zone has
a two-component oscillator [real and imaginary parts of $\varepsilon
(\vec{k}, t)]$, but since strain is real, the $\vec{k}$ and $-\vec{k}$
oscillators are images of each other (with real part the same, imaginary 
part
$\pi$ out of phase). Thus there are ${L_0}^2$ independent degrees of
freedom.

\bibitem{neutrality} For zero initial macroscopiic strain rates 
and strains, quasi-conservation is observed: the final
diagnostics
are $\langle \varepsilon_{\ell} \rangle \sim 10^{-3}$, with max/min 
$\varepsilon_{\ell}$ of order unity, (varying by a few tenths of a
percent over time near the end of a run).


\bibitem{GL} In Fourier space the effective  elastic constant
$A(k)\equiv F''_0 + U^c +K_0 k^2$ in the
Fourier version of (5.5) makes larger $k$ strains relax faster,
giving a weaker form of sequential-scale $BG$ relaxation.
Indeed when $K_0=0$, we get  blobs instead of
well-defined textures.


\bibitem{icomat02}T. Lookman, S.R. Shenoy, K.~{\O}. Rasmussen, A. Saxena, 
and A.R. Bishop, `On dynamics of models for ferroelastic transitions', 
Proceedings of ICOMAT'02, Helsinki, Finland, in press (2002).

\bibitem{delta} Taking the top right panel of Fig. 8 of energy per unit
cell $E_0 = -1.468$ as a reference, the fractional differences $\delta =
(E-E_0)/E_0$ of the two quasi-twin pictures  are $\delta = 0.0391$ and
0.0392. For yet another random-number seed a uniform state results,
with no domain walls (not shown). 


\bibitem{schmalian} J. Schmalian and P.G. Wolynes, Phys. Rev. Lett., 
{\bf 85}, 836 (2000)

\bibitem{Landau} L.D. Landau and E.M. Lifshitz, {\it Statistical Physics}
(Pergamon Press, Oxford, 1980); L.D. Landau, Phys. Zeit. d. Sow. Union {\bf 
8}, 113 (1935); {\bf 11}, 26, 545 (1937).


\end{references}
\end{document}